\documentstyle[12pt]{article}
\begin{document}
\title{\bf The Relation Between The Surface Brightness and The Diameter for 
Galactic Supernova Remnants}

\author{O. H. GUSEINOV$\sp1$
\thanks{huseyin@pascal.sci.akdeniz.edu.tr}
, A. ANKAY$\sp2$
\thanks{askin@gursey.gov.tr}
, A. SEZER$\sp1$
\thanks{sezer@pascal.sci.akdeniz.edu.tr} \\
and S. O. TAGIEVA$\sp3$
\thanks{physic@lan.ab.az} \\
\\ \\
{$\sp1$University of Akdeniz, Department of Physics,} \\ {Antalya,
Turkey} \\
{$\sp2$Bo\u{g}azi\c{c}i University - T\"{U}B\.{I}TAK, Feza G\"{u}rsey 
Institute,} \\ 
{81220 \c{C}engelk\"{o}y, \.{I}stanbul, Turkey} \\
{$\sp3$Academy of Science, Physics Institute Baku 370143,} \\ {Azerbaijan 
Republic}}

\date{}
\maketitle

\begin{abstract}
\noindent
In this work, we have constructed a relation between the surface 
brightness ($\Sigma$) and diameter (D) of Galactic C- and S-type supernova 
remnants (SNRs). In order to calibrate the $\Sigma$-D dependence, we have 
carefully examined some intrinsic (e.g. explosion energy) and extrinsic 
(e.g. density of the ambient medium) properties of the remnants and, 
taking into account also the distance values given in the literature, we 
have adopted distances for some of the SNRs which have relatively more 
reliable distance values. These calibrator SNRs are all C- and S-type 
SNRs, i.e. F-type SNRs (and S-type SNR Cas A which has an exceptionally 
high surface brightness) are excluded. The Sigma-D relation has 2 slopes 
with a turning point at D=36.5 pc: $\Sigma$(at 1 
GHz)=8.4$^{+19.5}_{-6.3}$$\times$10$^{-12}$ D$^{{-5.99}^{+0.38}_{-0.33}}$ 
Wm$^{-2}$Hz$^{-1}$ster$^{-1}$ 
(for $\Sigma$$\le$3.7$\times$10$^{-21}$ Wm$^{-2}$Hz$^{-1}$ster$^{-1}$ 
and D$\ge$36.5 pc) and $\Sigma$(at 1 GHz)=2.7$^{+2.1}_{-1.4}$$\times$ 
10$^{-17}$ D$^{{-2.47}^{+0.20}_{-0.16}}$ Wm$^{-2}$Hz$^{-1}$ster$^{-1}$
(for $\Sigma$$>$3.7$\times$10$^{-21}$ Wm$^{-2}$Hz$^{-1}$ster$^{-1}$ 
and D$<$36.5 pc). We discussed the theoretical basis for the $\Sigma$-D 
dependence and particularly the reasons for the change in slope of the 
relation were stated. Added to this, we have shown the dependence 
between the radio luminosity and the diameter which seems to have a 
slope close to zero up to about D=36.5 pc. We have also adopted distance 
and diameter values for all of the observed Galactic SNRs by examining 
all the available distance values presented in the literature together 
with the distances found from our $\Sigma$-D relation. 
\end{abstract}

\clearpage
\parindent=0.2in

\section{Introduction}
In the last 40 years it is known that there is a rough relation between 
the surface brightness value ($\Sigma$) in the radio band and the 
diameter (D) of Supernova remnants (SNRs) (Shklovsky 1960; Poveda \& 
Woltjer 1968; Clark \&Caswell 1976; Caswell \& Lerche 1979; Milne 1979; 
Lozinskaya 1981; Sakhibov \& Smirnov 1982; Allakhverdiev et al. 1983a; 
Allakhverdiev et al. 1983b; Huang \& Thaddeus 1985; Allakhverdiev et al. 
1986c; Green 1984; Li \& Wheeler 1984; Mills et al. 1984; Berkhuijsen 
1986; Case \& Bhattacharya 1998). 

If the whole SNR is bright and extends in a medium with roughly 
homogenious density then both its radio flux, F (mostly at 1 GHz), 
and its angular diameter, $\theta$ can be measured precisely. But in 
most of the cases, as the radiation coming from some parts of the SNR  
has low intensity the whole of the shell can not be observed (Green's  
catalog 2001; Caswell \& Lerche 1979; Milne 1979; Allakhverdiev et al. 
1986b). Also in 
some cases, as the SNR is projected onto a HII region its angular size 
and its flux can not be measured precisely (Weiler \& Panagia 1978; 
Israel 1980; Blandford \& Cowie 1982; Allakhverdiev et al. 1983b; Mills 
et al. 1984; Allakhverdiev et al. 1986a). As well known:  
\begin{equation} 
\Sigma\propto\frac{F}{\theta^2}\propto D^{-n}
\end{equation}
If only the bright part of 
the SNR is seen then the $\Sigma$ of the remnant will be overestimated, 
because the measured value of $\theta$ will be smaller than the real 
value of the angular size.  
What will be the effect of such an error on the distance value found from 
the $\Sigma$-D relation? From Eqn.1:
\begin{equation}
D\propto \frac{\theta^{2/n}}{F^{1/n}}
\end{equation}
Using Eqn.2 we can find a relation between distance (d), $\theta$ and F:
\begin{equation}
d=\frac{D}{\theta}\propto \frac{\theta^{(2-n)/n}}{F^{1/n}}
\end{equation}
If n=2 then overestimation (underestimation) of the $\Sigma$ ($\theta$) 
value has no effect on estimation of the distance value. If n $>$ 2 then  
overestimation (underestimation) of the $\Sigma$ ($\theta$) value leads 
to overestimation of the distance value and the degree of overestimating 
the distance value increases with n. If n $<$ 2 (which is not the case 
for the $\Sigma$-D relation as we will see below and this is known from 
various previously suggested $\Sigma$-D relations) the distance value will 
be underestimated in the case of underestimating (overestimating) the 
$\Sigma$ ($\theta$) value. An HII region projected onto the SNR may lead 
to overestimation of the flux value of the SNR and in such a case, from 
Eqn.3, there will be an underestimation of the distance value.  

Observationally, SNR distances are found in general using the shift value of 
21 cm HI line and galactic rotation models. By this method, for some of the 
SNRs (for which the 
radial velocity is large enough) distances can be found with percentage 
errors of about 30\%-50\% at best. The error increases  
in the vicinity of longitudes l=0$^o$ and l=180$^o$ and when a SNR is in 
or close to the galactic center direction 2 
values of distance, which are very different from each other, are found.  
If the SNR is found to be related with some other 
objects (which are located relatively close to the Sun) its distance can be 
determined more precisely, but in any case, 
error in the distance value is not smaller than about 20-30\%. So, in order 
to find distances of the SNRs, particularly the ones located at $>$3-4 
kpc, $\Sigma$-D relation 
is needed. 

In Figure 1 we see that the dispersion of the positions of the
calibrators (the SNRs which are used to calibrate the $\Sigma$-D 
relation) from the $\Sigma$-D relation is very high. The reasons for such 
high dispersions, in other words, the reasons for the $\Sigma$-D relation 
being not very reliable are discussed in detail by Allakhverdiev et al. 
(1986b). 

The increase in number of reliable calibrators do not decrease the 
dispersions much (Allakhverdiev et al. 1986b). The reasons for such 
dispersions are: \\
1) Explosion energies of supernovae (SNe) vary in a wide range, about 3 
orders of magnitude (e.g. kinetic energy of Crab SNR is $\sim$10$^{49}$ 
erg, Sollerman et al. 2000, whereas kinetic energy of Cas A SNR is 
$>$10$^{51}$ erg, Vink et al. 1998; Wright et al. 1999). So, it is 
important for constructing the $\Sigma$-D relation to examine the 
differences between the energies of calibrator SNRs. For example, SNR 
G11.2-0.3 has an explosion energy E$\sim$10$^{48}$-2.4$\times$10$^{49}$ 
erg (Bandiera et al. 1996), whereas SNR G320.4-1.2 has 
E$\sim$(1-2)$\times$10$^{51}$ erg (Gaensler et al. 1999b). For SNR 
G109.1-1.0 the energy may even be larger: E$\sim$10$^{51}$-10$^{52}$ 
(Morini et al. 1988). \\ 
2) Surface brightness of a SNR depends on density of the interstellar 
medium in which it expands as well as the kinetic energy of the expanding 
matter. The denser 
the medium is the higher the surface brightness will be under similar 
velocities (it must be noted 
that the lifetime of a SNR, in other words the time needed for the SNR's 
surface brightness to drop below a certain value, directly depends on the 
density of the medium and the SN explosion energy). \\
3) If there is a very active neutron star (NS) within the SNR which has a 
significant contribution to the SNR's energy, then the central part of 
the SNR can be much more bright. 

There are 231 SNRs observed up to date (Green, 2001). 155, 18 and 8 of 
them are S, C and F-type, respectively. For 50 of them type is 
either not known or not reliable. Naturally, number of SNRs 
increase as the SNRs are examined more precisely and as some new SNRs 
are found, but the increase in their number does not significantly change 
the ratio between different morphological types of SNRs. 
As S and C-type SNRs have very 
different characteristics compared to F-type SNRs and as there are not 
many F-type SNRs, $\Sigma$-D relation is constructed only for S and 
C-type SNRs and F-type remnants are not used as calibrators. As C-type 
SNRs are not different from S-type SNRs with respect to their energy and 
birth sites and because of the radiation coming from the plerionic part 
often being very low compared to the radiation coming from the shell, 
they can be used as calibrators for $\Sigma$-D relation. In general, for 
C-type SNRs the radiation coming from the shell is larger than the 
radiation coming from the plerionic part. But it must be noted that among 
the SNRs with D$>$20 pc the largest surface brightness values belong to 
the remnants W44, W28, RCW 89 and Milne 56 all of which are C-type SNRs. 

\section{Supernova Remnants as Calibrators}
We examined the SNRs given in Green (2001) and also some recently found 
SNRs with at least one observationally found distance value collecting 
radio, X-ray, and in some cases optical data of the SNRs. Among these 
SNRs we chose the ones with reliable distance values as calibrators to 
construct the $\Sigma$-D relation. The data of these SNRs which are 
essential for the $\Sigma$-D dependence and responsible for the 
deviations from the $\Sigma$-D dependence are given below (section 2.3). 
The calibrator SNRs are presented in the order of their galactic longitude 
(l) values and the calibrators with relatively more reliable distance 
values are presented before the calibrators with relatively less 
reliable distance values. All the calibrators are represented with 
asterisk (*) sign in Table 1. In Table 1, we represented the distances 
calculated using the $\Sigma$-D relation, the adopted distance values, 
the values of surface brightness and luminosity at 1 GHz for all of the 
Galactic SNRs.
While adopting the distances of the SNRs which are not calibrators, we 
have taken into account the distance values given in the literature, the 
data about the surrounding matter, the Galactic coordinates, and the 
$\Sigma$-D distances. The data and the references about the SNRs which 
were not adopted as calibrators, were taken from Tagieva (2002). In Table 
1, PSRs (P), X-ray pulsars (XP), pulsar wind nebula (PWN), anomalous 
X-ray pulsars (AXP), dim radio quiet neutron stars (D) connected to SNRs 
are also shown in the third column.   
\subsection{Abbreviations 
for SNR Data} 1. SNR type: S-Shell, C-Composite \\
2. Angular size for the shell and the plerionic part as a whole: $\theta$ 
(arcmin) \\
3. Radio spectral index for the shell and the plerionic part separately 
and for the whole SNR: $\alpha$ \\
4. Distance: d (kpc) \\
5. Column density of neutral hydrogen (HI): N$_{\hbox{\footnotesize {HI}}}$ 
cm$^{-2}$ \\ 
6. Interstellar optical absorption: A$_{\hbox{\footnotesize V}}$ (mag) \\
7. Radio flux at 1 GHz: F$_1$ (mJy) \\
8. Flux in X-ray band: F$_{\hbox{\footnotesize x}}$ erg/cm$^2$s \\
9. Temperature of plasma in the shell: kT (keV) \\
10. Velocity of shock front or expansion velocity: V (km/s) \\
11. Surface brightness: $\Sigma$ (Wm$^{-2}$Hz$^{-1}$sr$^{-1}$) \\
12. Radio luminosity at 1 GHz: (L$_{1}$ (Jy erg s$^{-1}$) \\
13. Luminosity in X-ray band: L$_{\hbox{\footnotesize x}}$ (erg/s) \\
14. For density of the SNR environment and types of clouds: 

a) Molecular cloud: MC 

b) Maser source (formed due to interaction between SNR and MC): MS 

c) Dust cloud: DC 

d) Neutral and ionized hydrogen clouds: HI and HII clouds

e) Number density of particles in front of SNR, in shell, in plerionic 
part, 

in different types of clouds and filaments: n (cm$^{-3}$) \\
15. Kinetic energy of SNR: E$_{\hbox{\footnotesize k}}$ (erg) \\
16. Explosion energy of SN: E (erg) \\
17. Age of SNR: t (kyr) \\
18. X-ray radiated mass: M$_{\hbox{\footnotesize x}}$ (M$_\odot$) \\
19. Ejected mass: M$_{\hbox{\footnotesize {Ej}}}$ (M$_\odot$) \\
20. Swept-up mass: M$_{\hbox{\footnotesize s}}$ (M$_\odot$) \\
21. Total mass: M (M$_{\odot}$) \\
22. Magnetic field in the shell: B (Gauss) 
\subsection{Abbreviations for Data of Point Sources Genetically Connected 
with SNRs}
1. Types of point sources:

a) Radio pulsar: PSR 

b) X-ray pulsar: XRP 

c) X-ray point source: XPS 

d) Neutron star: NS \\ 
2. Distance of point source from the geometrical center of SNR: 
$\beta$=$\Delta$$\theta$/$\theta$ \\
($\Delta$$\theta$: angular distance of point source from the geometrical 
center of SNR; $\theta$: average value of angular size of SNR) \\
3. Characteristic age of pulsar: $\tau$ (kyr) \\
4. Dispersion measure: DM pc cm$^{-3}$ 

Values of F$_1$, $\theta$ and $\alpha$ for SNRs were taken from Green 
(2001). 
Adopted distance values which were used in constructing the $\Sigma$-D 
diagram are also given below. 
\subsection{Calibrator Supernova Remnants}
{\bf SNR G4.5+6.8} \\
d=4.8$\pm$1.4 kpc [1], 5 kpc [2], 4.5$\pm$1 kpc [3], 4.1$\pm$0.9 kpc [4], 
d=4.4 kpc from the historical date and d=4.8-6.4 kpc from HI line 
measurements [5], d=4.8 kpc adopted. 

Kepler is about 500 pc above the galactic plane that it must be in a very 
low-density medium. So, its surface brightness must be much 
less than the surface brightness value corresponding to its diameter and 
its distance must be much less than the distance value found 
from the $\Sigma$-D relation. \\
$[1]$ Reynoso \& Goss 1999; [2] Borkowski et al. 1992; [3] Bandiera 1987; 
[4] Braun 1987; [5] Green 2001. \\
{\bf SNR G6.4-0.1} \\
d=3.5-4 kpc [1,20], d=2.5 kpc [2], d=2.5 kpc adopted. 

In the direction of this SNR, at 1.6 kpc, there is SGR OB1 association. 
Interstellar absorption, $A_V$, and $N_{HI}$ values of the stars, which 
are members of this association (l = 5$^o$.97 - 7$^o$.16, b = $-$0$^o$.48 
- +0$^o$.62), are $\sim$1$^m$ and $\sim$2$\times$10$^{21}$ cm$^{-2}$, 
respectively [3]. 

$N_{HI}$=(7-11)$\times$10$^{21}$ cm$^{-2}$, $N_{HI}$=3.5$\times$10$^{21}$ 
cm$^{-2}$ [4], $N_{HI}$=4.7$\times$10$^{21}$ cm$^{-2}$ [5,6]; 
E(B-V)=1-1.3$^m$ [4].

Since the SNR has $A_V$=3-4$^m$ and $N_{HI}$ $>$
3.5$\times$10$^{21}$ cm$^{-2}$ its distance is assumed to be 
considerably larger than the OB association's distance.

MS [10,13,16,17]; MC n=10$^5$ cm$^{-3}$ 
[9,10], n=2.5$\times$10$^4$ cm$^{-3}$ [11,12], n=2.5$\times$10$^4$ 
cm$^{-3}$ [13,15,16], n=30 cm$^{-3}$ (the average value in the shell) 
[14], n$_0$=0.1 cm$^{-3}$ [4], n$_0$=0.23 cm$^{-3}$ (the average value 
in front of the SNR) [6].

E=10$^{51}$ erg [4], E=4$\times$10$^{50}$ erg [6]; 
F$_x$=6.2$\times$10$^{-11}$ erg cm$^{-2}$s (0.5-2.4 
keV) [6]; t=2.5$\times$10$^3$ yr 
[4], t=(1-2.5)$\times$10$^4$ yr [6], t=(3.5-15)$\times$10$^4$ yr [7,8], 
t=6$\times$10$^4$ yr [14]; M$_x$=(19-26) M$_{\odot}$ [4]; B=0.2 mG (in 
the shell) [18]. 

If the SNR's explosion energy is 10$^{51}$ erg, the swept-up matter's 
mass must be about $\le$2$\times$10$^4$ M$_{\odot}$. In order to sweep-up 
such a mass the average density of the medium must be n$_0$$\le$5 
cm$^{-3}$. The SNR is in a very high-density medium [9,10].

The SNR interacts with the molecular cloud in the east of the SNR and this 
interaction has increased its radio and X-ray luminosity. There are many 
maser and HII regions in the field. The relativistic particles have 
energy of 2$\times$10$^{47}$ erg. It is C-type also in the X-ray band. 
The SNR is expanding in a very thick medium [19].

Since the medium is very dense, the values of the magnetic field and the 
explosion energy as well as the temperature and the X-ray luminosity 
must be large that this SNR should be very high above the $\Sigma$-D line. 
\\
$[1]$ Green 2001; [2] Sahibov 1983; [3] Ayd\i n et al. 1997; [4] Long et 
al. 1991; [5] Rho \& Petre 1998; [6] Rho et al. 1996; [7] Kaspi et al. 
1993; [8] Rowell et al. 2000; [9] Claussen et al. 1997; [10] Claussen et 
al. 1999a; [11] Wooten 1981; [12] Denoyer 1983; [13] Frail et al. 1994a; 
[14] Marsden et al. 1999; [15] Denoyer 1979a,b; [16] Frail et al. 1996; 
[17] Reach \& Rho 1998; [18] Koralesky et al. 1998a; [19] Dubner et al. 
2000; [20] Ankay \& Guseinov 1998. \\
{\bf SNR G31.9+0.0} \\
d=8.5 kpc (from 21cm HI line) [1], d=7.2 kpc [2], d=7.2 kpc (from 21cm HI 
line) [3,4], d=9 kpc [3,4], d=8.5 kpc adopted. 

$N_{HI}$=2.4$\times$10$^{22}$ cm$^{-2}$ (0.1-2.4 keV) [3,4]. 

The SNR is expanding in a dense medium [2]; MC [2]; 
n$_0$$\sim$5-10 cm$^{-3}$ (the average value in front of the SNR) [4], 
n$\sim$2$\times$10$^5$ cm$^{-3}$ (the average value behind the SNR, i.e. 
for the cloud) [5]; $M_{total}$$\cong$700 M$_{\odot}$ [4]. \\
$[1]$ Green 2001; [2] Frail et al. 1996; [3] Radhakrishnan et al. 1972; 
[4] Rho \& Petre 1996; [5] Reach \& Rho 1998. \\
{\bf SNR G34.7-0.4} \\
d=2.6 kpc [2], d=2.5 kpc [3], d=2.5 kpc (from 21cm HI line) [4], d=2.8 
kpc adopted.

$N_{HI}$=(1.6-2.1)$\times$10$^{22}$ cm$^{-2}$ [3].

No $H_2$O maser source [9]; MC [1,7,8]; n$_0$=6 cm$^{-3}$ (the average 
value in front of the SNR) [3], n$_0$=1 cm$^{-3}$ [3].

E$\cong$10$^{51}$ erg [3]; M=10$^3$ M$_{\odot}$ [3]; B=0.2 mG [10]. \\
\underline{Point Source PSR J1856+0113} \\
d=2.8 kpc [5,11], d=3.3 kpc [6]; $\tau$=2.03$\times$10$^4$ yr [11]; 
$\beta$=0.51 [13], $\beta$=0.6 [14].

A pulsar wind nebula (PWN) was found around this pulsar  
and this PWN is positionally coincident with the EGRET source 
[12]. \\
$[1]$ Giaconi et al. 1997; [2] Braun et al. 1989; [3] Cox et al. 1999; 
[4] Green 2001; [5] Kaspi 2000; [6] Taylor et al. 1996; [7] Frail et al. 
1996; [8] Denoyer 1979a,b; [9] Claussen et al. 1999; [10] Koralesky et 
al. 1998a; [11] Guseinov et al. 2002; [12] Roberts et al. 2001; [13] 
Lorimer et al. 1998; [14] Allakhverdiev et al. 1997. \\ 
{\bf SNR G54.4-0.3} \\ d=3.3 kpc [1], d=3.3 kpc adopted.  

MC and OB-associations are present [1]; n=30 cm$^{-3}$ for MC [1]; 
M$_s$$\cong$5$\times$10$^4$ M$_{\odot}$ [1]. \\
$[1]$ Junkes et al. 1992. \\
{\bf SNR G74.0-8.5} \\
d=0.8 kpc [1], d=0.7 kpc [2], d=1.4 kpc [4], d=460 pc [5], d=1.3$\pm$0.7 
kpc (kinematic distance) [3], d=440$^{+130}_{-100}$ pc (using the shock 
wave's velocity and proper motion) [6], d=0.8 kpc adopted.  

E(B-V)=0.08$^m$ [3] (As the absorption in this direction is small [7], this 
value is in accordance with a distance value of about 0.8-1 kpc). 

There is no open cluster in the direction of this SNR. In this direction, 
between 0.8-1.5 kpc, the reddening is almost constant [7]. \\
$[1]$ Minkowski 1958; [2] Braun et al. 1989; [3] Greidanus \& Strom 1992; 
[4] Sakhibov \& Smirnov 1983; [5] Braun \& Strom 1986; [6] Blair et al. 
1999; [7] Neckel \& Klare 1980. \\
{\bf SNR G78.2+2.1} \\
d=1.2 kpc [1], d=1.5 kpc [2,3,4,5,6], d=1.5 kpc adopted. 

MC [2]; n$_0$$\ge$4 cm$^{-3}$ [2]. 

The SNR is probably expanding inside a cavity [7]. 

$E_k$=1.7$\times$10$^{49}$ erg (for the shell) if d=1.5 kpc [2]; 
$M_x$=10$^2$ M$_{\odot}$ [7]. 

In the direction of the SNR (l = 76$^o$.8, b = 1$^o$.44), at d=1.37 kpc, 
there is Cygnus OB-association in which there are many massive stars   
[8]. \\ 
\underline{Point Source RX J2020.2+4026} [9] \\
$[1]$ Braun et al. 1989; [2] Landecker et al. 1980; [3] Green 1989; [4] 
Huang \& Thaddeus 1985; [5] Brazier et al. 1996; [6] Lorimer et al. 1998; 
[7] Lozinskaya et al. 2000; [8] Melnik \& Efremov 1995; [9] Brazier \& 
Johnston 1999. \\ 
{\bf SNR G109.1-1.0} \\
d=4 kpc [1], d=3.6 kpc [2], d=5 kpc [5,12], d=5.6 kpc [9], d=6 kpc [13],  
d=5 kpc adopted (since, the SNR is in a very dense medium and its 
explosion energy is high). \\

E(B-V)=(0.79-1.2)$^m$ [3]; $N_{HI}$=(8-10)$\times$10$^{21}$ cm$^{-2}$ 
[4], $N_{HI}$=4$\times$10$^{21}$ cm$^{-2}$ [5].

kT=0.17-0.56 keV [7]; F$_x$=7.8$\times$10$^{-11}$ erg cm$^{-2}$s (0.2-2.4 
keV) [7]; L$_x$=3.2$\times$10$^{37}$ erg/s [7]. 

MC [6,8]; n=20 cm$^{-3}$ (for the clouds) [3,11], n$_0$=0.25 
cm$^{-3}$ (the average value in front of the SNR) [5]. E=10$^{51}$ - 
10$^{52}$ erg [5]. \\
\underline{Point Source AXP 1E 2259+586} \\
d=5.6 kpc [9]; $N_{HI}$=9$\times$10$^{21}$ cm$^{-2}$ [7], 
$N_{HI}$=9.3$\times$10$^{22}$ cm$^{-2}$ (0.5-20 keV) [10]. \\
$[1]$ Green 1989; [2] Braun et al. 1989; [3] Fesen \& Hurford 1995; [4] 
Rho \& Petre 1993; [5] Morini et al. 1988; [6] Parmar et al. 1998; [7] 
Rho \& Petre 1997; [8] Tatematsu et al. 1990; [9] Hughes et al. 1984; 
[10] Patel et al. 2001; [11] Gotthelf \& Vasisht 1998; [12] Gaensler 
2000, [13] Hulleman et al. 2000. \\ 
{\bf SNR G114.3+0.3} \\ 
d=3.0 - 3.8 kpc [1], d=2.5-3 kpc [2,3], d=2.8 kpc adopted. 

n$_0$=0.1 cm$^{-3}$ (the average value for the medium) [4]. 

The SNR's shell expanded inside a HII region and has reached to the 
boundary of the HII region [4]. 

This SNR together with the SNRs G116.5+1.1 and G116.9+0.2 (CTB 1) are 
inside a supercavity [4].

In the direction of the SNR (l = 115$^o$.5, b = 0$^o$.25), at d=2.3 kpc, 
there is Cas 5 OB association [5]. \\
\underline{Point Source PSR J2337+6151} [6] \\
d=2.5 kpc (from 21cm HI line) [7], d=2.5 kpc [6], d=2.8 kpc [8]; 
$\tau$=4.07$\times$10$^4$ yr [8,11]; $\beta$=0.08 [9,10]. \\
$[1]$ Green 2001; [2] Reich \& Braunsfurth 1981; [3] Fesen et al. 1997; 
[4] Fich 1986; [5] Melnik \& Efremov 1995; [6] Lyne et al. 1996; [7] 
Becker et al. 1996; [8] Guseinov et al. 2002; [9] Lorimer et al. 1998; 
[10] Furst et al. 1993; [11] Taylor et al. 1996. \\ 
{\bf SNR G116.9+0.2} \\
d=3.1 kpc [1], d=2.3 kpc [2], d=2.8-4 kpc [3], d=3.5 kpc adopted. 

$N_{HI}$=7$\times$10$^{21}$ cm$^{-2}$ [5]; $A_V$=2.2-3.2 [6].

In the direction of this SNR, at 2.5 kpc, there is an O6-type star (HD/BD
108). For this star $N_{HI}$=3$\times$10$^{21}$ cm$^{-2}$ [4] that the
SNR's distance must be larger than the distance of this star.

In this direction, there is no star formation region beyond 3 kpc. If the 
distance of this SNR is $\sim$2.7 kpc, then it may be in the star 
formation region which include CAS OB2 and CAS OB5 associations. There is 
also a young open cluster (C2355+609, t=4$\times$10$^7$ yr) at about 3.7 
kpc in this direction [8]. 

For the stars in this direction, located at 1-4 kpc, $A_V$$\sim$2$^m$ and 
does not reach a value of 3$^m$ [7]. 

The SNR is expanding in a low-density supercavity which also include 
G114.3+0.3 and G116.5+1.1. The shock wave front does not have a regular 
but a discontinuous shape [6]. \\
$[1]$ Hailey \& Craig 1994; [2] Braun et al. 1989; [3] Green 2001; [4] 
Diplas \& Savage 1994; [5] Craig et al. 1997; [6] Fesen et al. 1997; [7] 
Neckel \& Klare 1980; [8] Lynga 1987. \\
{\bf SNR G119.5+10.2} \\
d=1.4 kpc [1,2,3,4,5], d=1.4 kpc adopted. 

$N_{HI}$=2.8$\times$10$^{21}$ cm$^{-2}$ [3], 
$N_{HI}$=(1.1-2.5)$\times$10$^{21}$ cm$^{-2}$ [2], 
$N_{HI}$=3.8$\times$10$^{21}$ cm$^{-2}$ [7]; $A_V$=1.3 [5]. 

MC [7]; n$\sim$1 cm$^{-3}$ (for the clouds in front of the 
SNR) [5], n$_0$$\sim$0.03 cm$^{-3}$ (the average value in front of the 
SNR) [5], n$_0$$\sim$0.02 cm$^{-3}$ (the average value in front of the 
SNR) [2].

E=3$\times$10$^{49}$ erg [2]; $M_s$=13 M$_{\odot}$ [2]; 
B=2.9$\times$10$^{-6}$ G [2].

OIII emission line ($\lambda$=5010 A, $\Delta$$\lambda$=28 A) is very 
strong as in SNRs G65.3+5.7 and G126.2+1.6 [8]. \\
\underline{Point Source RX J0007.0+7302} [6] \\
$[1]$ Pineault et al. 1993; [2] Seward et al. 1995; [3] Slane et al. 
1997; [4] Brazier et al. 1998; [5] Mavromatakis et al. 2000; [6] Brazier 
\& Johnston 1999; [7] Rho \& Petre 1998; [8] Fesen et al. 1981. \\
{\bf SNR G120.1+1.4} \\
d=2.2 kpc [1], d=2.3 kpc (from shock wave velocity model), d=4-5 kpc 
(from 21cm HI line) [2], d=4.6 kpc [3], d=3 kpc (from X-ray observations 
of Ginga satellite) [4], d=3.3 kpc adopted. 

$A_V$=2.1$\pm$0.5 [5]. 

In the direction of this SNR, there are Cas OB4 (at 2.7 kpc; l = 118$^o$ 
- 121$^o$.5, b = $-$2$^o$.7 - +2$^o$.3) and Cas 
OB7 (at 1.8 kpc; l = 121$^o$.8 - 124$^o$.2, b = $-$0$^o$.5 - +2$^o$.7) 
associations [8]. If the distance of this SNR is less than 3 kpc, i.e. 
if it is located close to the OB associations, its 
diameter must be less than 6 pc. Then, how can it be possible that the 
surface 
brightness of this young S-type SNR located in a dense medium is much less 
than the surface 
brightness value corresponding to its diameter? On the other hand, if 
Tycho's distance is 
close to 4 kpc, then the SNR may be in a low-density medium that its 
surface brightness can be such a low value [6,7]. \\
$[1]$ Albinson 1986; [2] Green 2001; [3] Schwarz et al. 1995; [4] Fink et 
al. 1994; [5] Chevalier et al 1980; [6] Reynoso et al. 1999; [7] Reynoso 
et al. 1997; [8] Garmany \& Stencel 1992. \\
{\bf SNR G132.7+1.3} \\
d=2.2$\pm$0.2 kpc (from 21cm HI line) [1], d=2.2 kpc [2], d=2.7 kpc [3], 
d=2.2 kpc (from optical data) [4], d=2.3 kpc adopted. 

The SNR's center is bright in X-ray [5].

$N_{HI}$$\sim$3$\times$10$^{21}$ cm$^{-2}$ [5], 
$N_{HI}$=6.9$\times$10$^{21}$ cm$^{-2}$ [6]; 
$N_{HI}$=4.3$\times$10$^{21}$ cm$^{-2}$ [10]; E(B-V)=0.71 [7].

The SNR interacts with the gas in the star formation region [1]. There 
are HII regions around the SNR [4]. HB3 is expanding in a dense medium 
[5]. 

kT=0.33 keV [10]; E=3.1$\times$10$^{50}$ erg [6]. \\
\underline{Point Source PSR J0215+6218} \\
d=2.3 kpc [8], d=3.2 kpc [9]; $\tau$=1.3$\times$10$^7$ yr [9]. \\
$[1]$ Routledge et al 1991; [2] Green 2001; [3] Braun et al. 1989; [4] 
Gray et al. 1999; [5] Rho et al. 1998; [6] Galas et al. 1980; [7] Fesen 
et al. 1995a; [8] Lorimer et al. 1998; [9] Guseinov et al. 2002; [10] Rho 
\& Petre 1998. \\ 
{\bf SNR G166.0+4.3} \\
d=4.5 kpc [2], d=3 kpc [3,4], d=3.8 kpc adopted.

The SNR's center is bright in X-ray [6,7,8].

In this direction, there is no identified star formation region at this 
distance, but the outer arm (Persei) of the galaxy is passing through. 
There is AUR OB2 association in the direction l = 172$^o$ - 174$^o$, b 
= $-$1$^o$.8 - +2$^o$.0 at 3.2 kpc. The distance of the SNR from the 
galactic plane at 4.5 kpc is 340 pc. Because of these the SNR is 
expected to be in a low-density medium. 

$N_{HI}$=2.9$\times$10$^{21}$ cm$^{-2}$ [5]. 

The SNR is expanding in a low-density cavity [8]. \\
$[1]$ Landecker et al. 1989; [2] Green 2001; [3] Allakhverdiev et al. 
1986b; [4] Braun et al. 1989; [5] Guo \& Burrows 1997; [6] Pineault et 
al. 1987; [7] Rho et al. 1994; [8] Fesen et al. 1997. \\
{\bf SNR G180.0-1.7} \\ 
d=0.8 kpc [1], d=1 kpc (by identifying some indications of the SNR in the 
spectrums of the stars in front of and behind the SNR) [2], d=1 kpc 
adopted (since distances of stars are used to find the distance of the 
SNR, this value is much more reliable). \\
$[1]$ Braun et al. 1989; [2] Phillips et al. 1981. \\
{\bf SNR G189.1+3.0} \\ 
d=1.5 kpc [1,2,3], d=1.5-2 kpc (from the interaction of the SNR with the 
MC [10], d=1.5 kpc adopted.

$N_{HI}$=(1-3)$\times$10$^{21}$ cm$^{-2}$ [3].

MC [5,6]; n=10-20 cm$^{-3}$ (the average value in front of the SNR) [8].

In the north-eastern part V$\cong$100 km/s and the density in front 
is n=10-1000 cm$^{-3}$, whereas in the southern part V$\cong$30 km/s and
the density in front is n=10$^4$ cm$^{-3}$ [8]. 

Not an H$_2$O maser source [7].

In the near infrared region the SNR's luminosity is 1.3$\times$10$^{36}$ 
erg/s [8].

In the direction of the SNR (l = 188$^o$.9, b = 3$^o$.44), at d=1.34-1.65 
kpc, there is Gem OB1 association in which there are 16 massive stars 
[11,12]. \\
B=500 $\mu$G [4]. \\ 
\underline{Point Source CXOU J061705.3+222127} [9] \\
In this region there is a point X-ray source [4]; 
$N_{HI}$=1.3$\times$10$^{21}$ cm$^{-2}$ [9,4].  

The radio flux at 327 MHz of the X-ray point source is not greater than 2 
mJy [9]. \\
$[1]$ Fesen 1984; [2] Allakhverdiev et al. 1986b; [3] Asoaka \& 
Aschenbach 1994; [4] Keohane et al. 1997; [5] Denoyer 1979a,b; [6] 
Frail et al. 1996; [7] Claussen et al. 1999; [8] Rho et al. 2001; [9] 
Olbert et al. 2001; [10] Green 2001; [11] Melnik \& Efremov 1995; [12] 
Blaha \& Humphreys 1989. \\
{\bf SNR G260.4-3.4} \\
d=1.5 kpc [1], d=2 kpc [2], d=2.2$\pm$0.3 kpc (from 21cm HI line) [3,4], 
d=1.9-2.5 kpc [5], d=1.3$^{+0.6}_{-0.8}$ kpc [6], d=2 kpc adopted.  

$N_{HI}$=(2-6)$\times$10$^{21}$ cm$^{-2}$ [7], 
$N_{HI}$=(2.9-4.7)$\times$10$^{21}$ cm$^{-2}$ [5,8], 
$N_{HI}$=1.4$\times$10$^{21}$ cm$^{-2}$ [9].

There is a B0.7 Ib type star (HD 69882; l = 259$^o$.5, b = $-$3$^o$.9, 
d=2.1 kpc) in the same region of Puppis A and for this star 
$N_{HI}$=1.6$\times$10$^{21}$ cm$^{-2}$ [10].

Puppis A is in the direction of Vela X and, like the star formation 
regions in this part of the galaxy, it is below the geometrical plane of 
the galaxy.

Puppis A is not exactly in the direction of the star formation regions. 
Distances of the OB associations in the star formation region do not exceed 
1.5-1.8 kpc [11,12]. 

On the other hand, it is seen from the distribution of neutral hydrogen 
(HI) in the galaxy that the cold clouds in the direction of Puppis A are 
in general nearer than 1.5-1.8 kpc [1].

Diameter of this SNR has reached to 32 pc and it has gone out of the HII 
region it was in. Eastern part of the remnant is interacting with the HI 
cloud [3]. 

There are OH clouds in front of the SNR, but no sign of an interaction 
between the SNR and the clouds has been found [6]. 

n$_0$$\cong$0.4-0.5 cm$^{-3}$ (the average value in front of the SNR) 
[3], n=100 cm$^{-3}$ (for X-ray emitting region) [1], n=10-1000   
cm$^{-3}$ (for clouds) [3], n$_0$=3 cm$^{-3}$ (the average value in front 
of the SNR) [7], n$_0$=1 cm$^{-3}$ [9].

N, O and Ne are abundant in the SNR [2]. \\
\underline{Point Source RX J0822-4300} [13] \\
$N_{HI}$=(4-8)$\times$10$^{21}$ cm$^{-2}$ [2].

The region $<$30$''$ around the pulsar in the SNR has been examined. 
The upper limit of the radio luminosity of a possible pulsar-powered 
nebula is 3 orders of magnitude less than what would be expected if RX 
J0822-4300 was an energetic young radio pulsar beaming away from us. RX 
J0822-4300 has some properties which are very different compared to most 
of the young pulsars' properties [14]. \\
$[1]$ Braun et al. 1989; [2] Petre et al. 1996; [3] Reynoso et al. 1995; 
[4] Green 2001; [5] Zavlin et al. 1999; [6] Woermann et al. 2000; [7] 
Winkler et al. 1981a,b; [8] Blair et al. 1995; [9] Berthiaume et al. 1994; 
[10] Diplas \& Savage 1994; [11] Melnik \& Efremov 1995; [12] Humphreys 
1978; [13] Pavlov et al. 1999; [14] Gaensler et al. 2000b. \\
{\bf SNR G263.9-3.3} \\ 
Recent distance estimates of Vela are as follows: d=0.25 kpc [1], 
d=0.25$\pm$0.03 kpc [2], d$\sim$0.28 kpc [3], d=0.25,0.3 kpc [14].

The stars which are in front of, behind, and interacting with Vela have  
been identified [4]. The distance of Vela is given as 250$\pm$30 pc in 
[4].

In estimating 
the distance one should also consider that Vela SNR expands in a dense 
environment. Its magnetic field is B$\sim$6$\times$10$^{-5}$ G [5]
and its explosion energy is (1-2)$\times$10$^{51}$ erg [4].
These values have large errors, but the values themselves 
are also large. If we take all of these values into account, then, in the 
$\Sigma$-D diagram, it is not acceptable to put Vela at the same position 
with SNR G327.6+14.6 (remnant of type Ia supernova explosion at 500 pc 
above the galactic plane [6]) which expands in a low-density medium. 

In the direction of Vela remnant, none of the young open clusters nor OB 
associations have distances as small as 0.25 kpc [7],[8],[9]. 
Among the open clusters in the direction of Vela SNR (none of them has a
distance value as small as 0.25 kpc) Pismis 4 (l = 262$^o$.7, b =
$-$2$^o$.4) and Pismis 6 (l = 264$^o$.8, b = $-$2$^o$.9), which have the
most precise distance values and are exactly in the direction of Vela
SNR, are located at 0.6 kpc and 1.6 kpc, respectively [1].
Since progenitors of SNRs (and 
pulsars) are massive stars one would expect Vela to be closer to the star 
formation region instead of a distance value of 0.25 kpc. 

If the distance value of 0.45 kpc (which is close to the previous 
distance estimation of 0.5 kpc [10]) is 
adopted for Vela PSR, then the average electron density along the line of 
sight will be n$_e$=0.153 cm$^{-3}$. The pulsar with the second largest 
n$_e$ value ($\sim$0.113 cm$^{-3}$) is for PSR J1302-6350 (l = 
304$^o$.2, b = $-$0$^o$.9, d=1.3 kpc, with a Be type companion, variable 
wind in the environment) and the third largest n$_e$ value (0.107 
cm$^{-3}$) belongs to PSR J1644-4569 (l = 339$^o$.2, b = $-$0$^o$.2). 
Since the flux of PSR J1644-4569 at 1400 MHz is larger than any other 
known pulsars' flux value at the same frequency, we can estimate its 
distance as not more than 4.5 kpc. Average value of n$_e$ for the rest of 
pulsars is about 0.04 cm$^{-3}$ [13]. So, it is not possible to accept a 
distance value of 0.25 kpc for Vela pulsar and Vela SNR. All we could do 
is to reduce our initial distance estimate of 0.45 kpc to at most 0.4 
kpc. 

V=170 km/s [11]; kT=0.086-0.17 keV [12]; A$_V$=0.56 [11]; 
E=(1-2)$\times$10$^{51}$ erg [4]; B$\sim$6$\times$10$^{-5}$ G [5]. 

From the above discussion d=0.45 kpc adopted. \\
\underline{Point Source PSR J0835-4510} \\
DM=68.2 cm$^{-3}$pc; d=0.5 kpc [15]; $\tau$=1.1$\times$10$^4$ yr [13,15]; 
$\beta$=0.29 [16], $\beta$=0.3 [17]. \\
$[1]$ \"{O}gelman et al. 1989; [2] Cha et al. 1999; [3] Bocchino et al. 
1999; [4] Danks 2000; [5] de Jager et al. 1996; [6] Hamilton et al. 1997; 
[7] Efremov 1989; [8] Berdnikov \& Efremov 1993; [9] Ayd\i n et al. 
1997; [10] Green 2000; [11] Raymond et al. 1997; [12] Kahn et al. 
1985; [13] Guseinov et al. 2002; [14] Green 2001; [15] Taylor et al. 
1996; [16] Lorimer et al. 1998; [17] Allakhverdiev et al. 1997. \\ 
{\bf SNR G296.5+10.0} \\
d=1.5 kpc [1,2,3], d=1-2 kpc [4], d=2.1 kpc [5], d=1.8 kpc adopted. 

$N_{HI}$=1.4$\times$10$^{21}$ cm$^{-2}$ (d=1-2 kpc) [6], 
$N_{HI}$=(1.1-1.6)$\times$10$^{21}$ cm$^{-2}$ (d=1-2 kpc) [5], 
$N_{HI}$=4$\times$10$^{20}$ cm$^{-2}$ (0.1-10 keV) [7], 
$N_{HI}$=4$\times$10$^{20}$ cm$^{-2}$ [2]; $A_V$(r)=0.5$^m$ (d=1-2 kpc) 
[8].

n$_0$=0.2 cm$^{-3}$ (the average value in front of the SNR) [7]; 
E=2$\times$10$^{50}$ erg [7], E=6$\times$10$^{50}$ erg (d=1-2 kpc) [4], 
E$>$2$\times$10$^{49}$ erg [5].

Mass of the neutral hydrogen in the SNR is more than 1900 M$_{\odot}$ 
(d=1-2 kpc) [5]. \\
\underline{Point Source 1E 1207.4-5209} [2,5] \\
There is a hole in the HI clouds, at exactly the center of the SNR and 
the neutron star is located at this position that it is genetically 
connected with the SNR [5]. \\
$[1]$ Kaspi et al. 1996; [2] Mereghetti et al. 1996; [3] Zavlin et al. 
2000; [4] Roger et al. 1988; [5] Giacani et al. 2000; [6] Kellett et al. 
1987; [7] Vasisht et al. 1997; [8] Ruiz 1983. \\
{\bf SNR G315.4-2.3} \\ 
d=2.8 kpc [1,2], d=2.5 kpc [3,4], d=2.8 kpc (kinematic) [6], d=2.5 kpc 
[10], d=2.5 kpc (the SNR is located in an OB-association) [13], d=2.7 kpc 
adopted. 

$N_{HI}$=(1-4)$\times$10$^{21}$ cm$^{-2}$ [2,5], 
$N_{HI}$=3$\times$10$^{21}$ cm$^{-2}$ [10].

n$_0$=0.2 cm$^{-3}$ (the average value in front of the SNR), n=10 
cm$^{-3}$ (for the clouds) [6], n=0.2 cm$^{-3}$ (the average value in 
front of the SNR) [12], n$_0$=0.3 cm$^{-3}$ (the average value in front of 
the SNR) [13]; E=6.6$\times$10$^{50}$ erg [6].

The part of the remnant, which is bright in radio, is also bright in 
X-ray [7]. X-ray synchrotron radiation has been observed [7,8,9].

The abundances of O, Ne, Mg, and Si are more than the abundance of Fe that 
the SNR was formed due to a type-II supernova [10]. 

The SNR's morphology looks like the morphology of Tycho. The SNR has 
expanded inside a cavity and now it seems that it is expanding in the 
boundary of the cavity. This leads to a rapid drop in SNR's expansion 
velocity [11]. 

In the direction of the SNR (l = 315$^o$.5, b = $-$2$^o$.75), at d=2.5 
kpc, there is Cir OB1 association [14]. \\
$[1]$ Greidanus \& Strom 1992; [2] Petruk 1999; [3] Braun et al. 1989; 
[4] Green 2001; [5] Nugent et al. 1984; [6] Rosado et al. 1996; [7] 
Borkowsky 2001; [8] Allen et al. 1998; [9] Asvarov et al. 1990; 
[10] Bamba et al. 2000; [11] Dickel et al. 2001; [12] Long \& Blair 
1990; [13] Borkowsky et al. 2001; [14] Blaha \& Humphreys 1989. \\
{\bf SNR G320.4-1.2} \\ 
d=3.6 kpc [1], d=4kpc [2], d=5.2 kpc (from 21cm HI line) [3], d=5.2 kpc 
[4], d=4 kpc [11], d= 4.2 kpc adopted. 

$N_{HI}$=9$\times$10$^{21}$ cm$^{-2}$ [5], $N_{HI}$=9.5$\times$10$^{21}$ 
cm$^{-2}$ [6], $N_{HI}$=6$\times$10$^{21}$ cm$^{-2}$ [7].

n=100 cm$^{-3}$ (in the X-ray emitted part) [1]; 
E=(1-2)$\times$10$^{51}$ erg [4]; M=28 M$_{\odot}$ (possible) [10]; 
B$\sim$8$\times$10$^{-6}$ G (in the plerionic part) [10]. \\
\underline{Point Source PSR J1513-5908} [8,9] \\
d=4.2 kpc [9,12]; $N_{HI}$=5.9$\times$10$^{21}$ cm$^{-2}$ [11]; 
$\tau$=1.55$\times$10$^3$ yr [9,12]; $\beta$=0.24 [13]; a jet has been 
observed [11]. \\
$[1]$ Braun et al. 1989; [2] Allakhverdiev et al. 1986b; [3] Green 2001; 
[4] Gaensler et al. 1999a; [5] Seward et al. 1984; [6] Greiveldinger et 
al. 1995; [7] Trussoni et al. 1996; [8] Allakhverdiev et al. 1997; [9] 
Taylor et al. 1996; [10] du Plessis et al. 1995; [11] Tamura et al. 
1996; [12] Guseinov et al. 2002; [13] Lorimer et al. 1998. \\ 
{\bf SNR G327.6+14.6} \\ 
d=0.7$\pm$0.1 kpc (from Sedov's model) [1], d=1.5-2.5 kpc [8], d=1.7-3.1 
kpc [4], d=2 kpc [7], d=1.8$\pm$0.3 kpc [2,3,4], d=2 kpc adopted. 

$N_{HI}$=(3.9-5.7)$\times$10$^{20}$ cm$^{-2}$ [1], 
$N_{HI}$=1.8$\times$10$^{21}$ cm$^{-2}$ [9]; $A_V$=0.31 [10].

n$_0$=0.4 cm$^{-3}$ (ambient density) [1], n$_0$=0.1 cm$^{-3}$ (in 
front of the SNR) [12], n$_0$$\sim$0.02 cm$^{-3}$ [11].

E$>$4.4$\times$10$^{49}$ erg [1], E=10$^{51}$ erg [12]. 

B=(6-10)$\times$10$^{-6}$ G [9], B=(3-6)$\times$10$^{-6}$ G [12].

Using the age (t$\sim$1000 yr) and the expansion velocity (16600 km/s) of 
this SNR its diameter is found to be $\sim$17 pc. As the 
angular diameter of SN 1006 is 
30$'$ the lower limit for its distance value should be 1.9 kpc [5,6,7]. \\
$[1]$ Willingale et al. 1996; [2] Long et al. 1988; [3] Roger et al. 
1988; [4] Green 2001; [5] Fesen 1988; [6] Wu et al. 1993; [7] Winkler \& 
Long 1997; [8] Schaefer 1996; [9] Koyama et al. 1995; [10] Laming et al. 
1996; [11] Krishner et al. 1987; [12] Reynolds 1996. \\
{\bf SNR G332.4-0.4} \\ 
d=4 kpc [9], d=3.7 kpc adopted.

$A_V$=4.5$^m$ [1]. This value is comparable with the average $A_V$ value 
of 6.3$^m$ [3] for the stars at d=3.4 kpc [2] in this direction. The 
distance has mostly been assumed to be 3.3 kpc [4,5,6,7]. 

In the direction of RCW 103, there is R103 cluster located at 4 kpc [8]. 

MC [15]; n=1000 cm$^{-3}$ (for the clouds) [8], n$\ge$1000 cm$^{-3}$ 
(for the MC) 
[14]. Density of the gas clouds behind the shock wave is relatively low 
(n$_e$$\sim$10$^3$ cm$^{-3}$) [11]. 

The SNR's angular radius has increased 1$^{''}$.8$\pm$0$^{''}$.2 in 25 
years [13]. 

$N_{HI}$=6.8$\times$10$^{21}$ cm$^{-2}$ [10] value shows the possibility 
that RCW 103 might be related with the cluster R103. If the SNR is 
in the same region with the cluster, then it is possible that the remnant 
is expanding in a dense medium. 

The SNR has an approximately spherical shape. It is bright and has a thick 
shell. Northern part of the shell interacts with the molecular cloud. The 
SNR can be assumed to have formed due to a supernova explosion 1000 years 
ago [12]. \\
$[1]$ Oliva et al. 1990; [2] Caswell et al. 1975; [3] Neckel \& Klare 
1980; [4] Tuohy \& Garmire 1980; [5] Gotthelf et al. 1997; [6] Kaspi et 
al. 1996; [7] Green 2001; [8] Braun et al. 1989; [9] Allakhverdiev et al. 
1986b; [10] Gotthelf et al. 1999b; [11] Oliva et al. 1999; [12] Dickel et 
al. 1996; [13] Carter et al. 1997; [14] Meaburn \& Allan 1986; [15] Frail 
et al. 1996. \\
{\bf SNR G5.4-1.2} \\ 
d=4.5 kpc (from 21cm HI line) [1], d$\cong$5 kpc [2], d$>$4.3 kpc (from 
21cm HI line) [3], d=4.5 kpc adopted.

n$_0$$>$3$\times$10$^{-3}$ cm$^{-3}$ (in front of the SNR) [4]. \\
\underline{PSR J1801-2451} \\
SNR - PSR J1801-2451 connection [1,4]; d=4.5 kpc [6], d=4.4 kpc [5]; 
$\tau$=1.5$\times$10$^4$ yr [5,6]; $\beta$=0.8 [7], $\beta$$\sim$1 [8]. \\
$[1]$ Frail et al. 1994b; [2] Caswell et al. 1987; [3] Green 2001; [4] 
Frail \& Kulkarni 1991; [5] Taylor et al. 1996; [6] Guseinov et al. 2002; 
[7] Allakhverdiev et al. 1997; [8] Lorimer et al. 1998. \\ 
{\bf SNR G11.2-0.3} \\
d=5 kpc (from 21cm HI line) [1], d=5 kpc adopted. 

$N_{HI}$$\cong$10$^{22}$ cm$^{-2}$ [3], $N_{HI}$=1.38$\times$10$^{22}$ 
cm$^{-2}$ [4]; E$\sim$10$^{48}$ - 2.4$\times$10$^{49}$ erg [2]; 
t$\sim$2 kyr [7]. \\ 
\underline{Point Source AX J1811-1926} \\
d=5 kpc [5]; $\tau$=2.4$\times$10$^4$ yr [6].

If the pulsar's real age is 2.4$\times$10$^4$ yr and if the SNR is 
a historical one, then the two age values contradict with each other. The 
possibility of a genetic connection between the radio-quiet pulsar and the 
SNR is examined in [7]. \\
$[1]$ Green 2001; [2] Bandiera et al. 1996; [3] Reynolds et al. 1994; [4] 
Vasisht et al. 1996; [5] Kaspi 2000; [6] Torii et al. 1999; [7] Roberts 
et al. 2000. \\
{\bf SNR G43.3-0.2} \\
d=8.5 kpc [1], d=12.5-14 kpc (from 21cm HI line) [2], d=9 kpc adopted. 

$N_{HI}$=4$\times$10$^{22}$ cm$^{-2}$ (0.5-10 keV) [3].

The medium around the SNR seems to be highly dense from X-ray 
observations [4]. \\
$[1]$ Braun et al. 1989; [2] Green 2001; [3] Fujimoto et al. 1995; [4] 
Hwang et al. 1999. \\
{\bf SNR G69.0+2.7} \\
d=1.3 kpc [1], d=2 kpc adopted; E=10$^{51}$ erg. \\
\underline{Point Source J1952+3252 (radio and X-ray pulsar)} [4] \\
d=2.5 kpc [2,3,4], d=2 kpc [5]; $N_{HI}$=3$\times$10$^{21}$ 
cm$^{-2}$ [2,3]; $\tau$=1.07$\times$10$^5$ yr [4,5]; $\beta$=0.14 [6], 
$\beta$=0.15 [7]. \\
$[1]$ Braun et al. 1989; [2] Safi-Harb et al. 1995; [3] \"{O}gelman \& 
Buccheri 1987; [4] Taylor et al. 1996; [5] Guseinov et al. 200; [6] 
Lorimer et al. 1998; [7] Allakhverdiev et al. 1997. \\ 
{\bf SNR G89.0+4.7} \\
d=0.8 kpc (from the connection of the remnant with the association Cyg 
OB7) [1,2], d=0.9 kpc adopted. \\
$[1]$ Huang \& Thaddeus 1986; [2] Tatematsu et al. 1990. \\
{\bf SNR G93.3+6.9} \\
d=3.8 kpc [1], d=2.5 kpc (from 21cm HI line) [2], d=3.8 kpc adopted 
(z=420 pc, D=26 pc). 

$N_{HI}$=5.7$\times$10$^{21}$ cm$^{-2}$ [2];  
E=3.9$\times$10$^{50}$ erg [1].

The type-Ia supernova has exploded 5000 years ago [1]. 

As this SNR is in a very low-density medium, its position on the 
$\Sigma$-D diagram should be well below the $\Sigma$-D line. \\
$[1]$ Landecker et al. 1999; [2] Green 2001. \\
{\bf SNR G156.2+5.7} \\
The data from ROSAT X-ray satellite were examined using Sedov model. 
Taking the results found from this model and 
$N_{HI}$=8.8$\times$10$^{20}$ cm$^{-2}$ value into account d=3 kpc  
[1,3]. d=1.3 kpc [4]; D=100 pc if d=3 kpc [4].

$N_{HI}$=9$\times$10$^{20}$ cm$^{-2}$ [2,3].

There is no star formation region in the direction of this SNR, but there 
is an open cluster for which the distance is not known. Both of the 
distance values correspond to z$>$100 pc (for d=1.3 kpc z=130 pc). The 
facts that there is no star formation region in this direction and that 
z$>$130 pc show that the remnant is in a low-density medium. So, the 
SNR's diameter (and distance) is not expected to be larger than the 
diameter value corresponding to its surface brightness in the $\Sigma$-D 
diagram. The diameter might be less than the diameter corresponding 
to its surface brightness. As a result, d=2 kpc adopted (z=200 pc). \\
$[1]$ Pfeffermann et al. 1991; [2] Yamauchi et al. 1993; [3] Reich et al. 
1992; [4] Yamauchi et al. 1999. \\
{\bf SNR G205.5+0.5} \\
d=0.8 kpc (from optical data), d=1.6 kpc (from radio data) [1], d=1.6 kpc 
[2], d=1 kpc adopted. \\
$[1]$ Green 2001; [2] Odegard 1986. \\
%
{\bf SNR G111.7-2.1} \\
d=2.8 kpc [1], d=3.4 kpc (by examining the SNR's dynamics) [2].

Cas A has no projection on the OB-associations and none of 
these OB-associations, which are in the directions close to the SNR, has a 
distance $>$3 kpc [3,4]. 

$N_{HI}$=1.2$\times$10$^{21}$ cm$^{-2}$ [5]; 
E$\sim$10$^{51}$ erg [6,10].

$M_s$+$M_{Ej}$=7-12 M$_{\odot}$ [5], $M_{Ej}$$\sim$4 M$_{\odot}$, 
$M_s$+$M_{Ej}$$\sim$12 M$_{\odot}$ [6].

Since Cas A was born because of a massive star's explosion in a dense 
interstellar medium, its distance was adopted as 3 kpc. This SNR was not 
used as a calibrator. \\
\underline{Point Source RQNS CXO J2323+5848} [7] \\
d=3.4 kpc [9]; $N_{HI}$=1.1$\times$10$^{22}$ cm$^{-2}$ [8]; $A_V$=5$^m$ 
[9]. \\
$[1]$ Green 2001; [2] Reed et al. 1995; [3] Humphreys 1978; [4] Garmany 
\& Stencel 1992; [5] Favata et al. 1997; [6] Vink et al. 1998; [7] 
Brazier \& Johnston 1999; [8] Chakrabarty et al. 2001; [9] Kaplan et al. 
2001; [10] Wright et al. 1999. \\ 
{\bf SNR G130.7+3.1} \\
d=3.2 kpc [1,2,3,4,5], d=2.2 kpc [6], d=2.6 kpc [7], d=3 kpc adopted.

The SNR is plerion in X-ray. 

$N_{HI}$=1.8$\times$10$^{21}$ cm$^{-2}$ (2-10 keV) [1,8], 
$N_{HI}$=2$\times$10$^{21}$ cm$^{-2}$ (0.5-4.5 keV) [9], 
$N_{HI}$=(3-4)$\times$10$^{21}$ cm$^{-2}$ [10], 
$N_{HI}$=3$\times$10$^{21}$ cm$^{-2}$ [3]. 

$A_V$=1.3$\pm$0.2 (if this value is correct, then, according to [11] 
d$<$1 kpc); B=3$\times$10$^{-3}$ G [10]. \\
\underline{X and radio PSR J0205+6449} [12,13]  \\
d=3.2 kpc [14]; $\tau$=5.5$\times$10$^3$ yr [14]. \\
$[1]$ Roberts et al. 1993; [2] Frail \& Moffett 1993; [3] Helfand et al. 
1995; [4] Lorimer et al. 1998; [5] Green 2001; [6] Braun et al. 1989; [7] 
Allakhverdiev et al. 1986b; [8] Davelaar et al. 1986; [9] Becker et al. 
1982; [10] Torii et al. 2000; [11] Neckel \& Klare 1980; [12] Murray et 
al. 2002; [13] Camilo et al. 2002; [14] Guseinov et al. 2002. \\ 
{\bf SNR G184.6-5.8} \\
d=2 kpc [1], d=2 kpc adopted.

$N_{HI}$=3$\times$10$^{21}$ cm$^{-2}$ [3]; E(B-V)=0.52$^m$ [3]; R=3.1$^m$ 
[3]. 

E$\sim$10$^{49}$ erg; L$_x$=10$^{37}$ erg/s (2-10 keV) [5], 
L$_x$=2.5$\times$10$^{37}$ erg/s [6].

$\Sigma$$<$4.3$\times$10$^{22}$ Wm$^{-2}$Hz$^{-1}$ster$^{-1}$ (if 
the SNR has a shell) [2].

The CIV ion's line at $\lambda$=1550 A shows that there may be a shell 
near Crab moving with a velocity of 2500 km/s. 
E=1.5$\times$10$^{49}$ erg [3]. \\
\underline{Point Source J0534+2200} \\
d=2 kpc; $\beta$$\sim$0.1 [4]; $\tau$=1.26$\times$10$^3$ yr [7,8]. \\
$[1]$ Green 2001; [2] Frail et al. 1995; [3] Sollerman et al. 2000; [4] 
Lorimer et al. 1998; [5] Davelaar et al. 1986; [6] Becker et al. 1982; 
[7] Taylor et al. 1996; [8] Guseinov et al. 2002. \\ 

After examining the data of these SNRs and the point sources in them, we 
adopted distance values for the SNRs.

Among the 34 SNRs given above the first 23 ones have the most 
reliable distance values. For the next 8 SNRs the distance 
values are relatively less reliable. The last 3 SNRs presented above (Cas 
A (G111.7-2.1), Crab (G184.6-5.8), and SN 1181 (G130.7+3.1)) were not 
chosen as calibrators; the surface brightness value of Cas A is 
extraordinarily high and, Crab and SN 1181 are F-type SNRs. These SNRs
are shown on the $\Sigma$-D diagram, because their distances are known 
precisely and their diameters are small. We included them in the $\Sigma$-D 
diagram just to see their positions on the diagram.   

\section{Constructing $\Sigma$-D Relation Using Calibrators}
Using the 31 SNRs given above (for which reliable distance values 
were determined) we constructed the $\Sigma$-D relation (Fig. 1). 
In the $\Sigma$-D diagram, F-type historical SNRs Crab and SN 1181, 
for which the distances are well known, are shown as cross (x) signs. As 
mentioned above, for F-type SNRs $\Sigma$-D relation can not be used, 
though the positions of Crab and SN 1181 are shown on the $\Sigma$-D 
diagram just to see deviations of their positions from the $\Sigma$-D 
relation. 

SNR Cas A has the highest energy among the Galactic SNRs which were 
formed by SN explosion in the last two thousand years  (for all of 
these SNRs dates of explosion are known). The massive shell of Cas A is 
expanding through a 
dense medium, so that, the magnetic field behind the shock 
wave is more intense and the situation is more convenient in order to 
accelarate the electrons (independent of the accelaration mechanism). 
Because of these reasons Cas A, for which the distance is well known, 
deviates from the $\Sigma$-D line more than the other SNRs 
($\Sigma$ $<$ 10$^{-18}$ Wm$^{-2}$Hz$^{-1}$ster$^{-1}$) do. In our 
galaxy, this SNR has the highest $\Sigma$ (and also luminosity) value and 
it's very high $\Sigma$ value makes it unique. 

As seen from Fig.1, the 
historical SNRs G4.5+6.8 (Kepler), G332.4-0.4 (RCW 103) and G327.6+14.6 
(SN 1006) are located below the $\Sigma$-D line. Since, Kepler and SN 
1006 are very far away from the Galactic plane they are expanding in a 
low-density medium and because of this they have surface brightness 
values much less than the surface brightness values corresponding to 
their diameters. SNRs G43.3-0.2 (W49B), G6.4-0.1 (W28), G320.4-1.2 
(RCW89) and G132.7-1.3 (HB3) are located above the $\Sigma$-D line.
 
For the calibrator SNRs shown in Fig.1 two $\Sigma$-D relations with 
different slopes were constructed; one for the SNRs having $\Sigma$ $\le$ 
3.7$\times$10$^{-21}$ 
Wm$^{-2}$Hz$^{-1}$ster$^{-1}$ (D $\ge$ 36.5 pc) and the other for the SNRs 
with $\Sigma$ $>$ 3.7$\times$10$^{-21}$ Wm$^{-2}$Hz$^{-1}$ster$^{-1}$ (D 
$<$ 36.5 pc):  
\begin{equation}
\Sigma=8.4^{+19.5}_{-6.3} \times 10^{-12} D^{{-5.99}^{+0.38}_{-0.33}} 
\hspace{0.3cm} W m^{-2} Hz^{-1} ster^{-1}
\end{equation}
and
\begin{equation}
\Sigma=2.7^{+2.1}_{-1.4} \times 10^{-17} D^{{-2.47}^{+0.20}_{-0.16}} 
\hspace{0.3cm} W m^{-2} Hz^{-1} ster^{-1}
\end{equation} 
Since we have used the flux values at 1 GHz given in Green (2001) these 
equations are valid only for 1 GHz frequency. 

In Fig.2, the relation between radio luminosity (at 1 GHz) and diameter 
values of the calibrator SNRs shown in Fig.1 are given. Similar to the 
$\Sigma$-D relation, we found 2 equations for L-D relation with an 
intersection at L=5300 Jy kpc$^2$, D=36.5 pc; one for the SNRs having L $>$ 
5300 Jy kpc$^2$, D $<$ 36.5 pc and the other for the SNRs having L $\le$ 
5300 Jy kpc$^2$, D $\ge$ 36.5 pc: 
\begin{equation}
\L=2.45 \times 10^4 D^{-0.43} \hspace{0.3cm} Jy \hspace{0.1cm} kpc^2
\end{equation}
and
\begin{equation}
\L=5.38 \times 10^9 D^{-3.84} \hspace{0.3cm} Jy \hspace{0.1cm} kpc^2
\end{equation}

These dependences are not as reliable as the $\Sigma$-D 
dependences given above (Eqns. 4 and 5) that we did not give their 
errors which are very large. Similar to 
the $\Sigma$-D diagram, in Fig.2 Cas A and Crab have very high radio 
luminosities. Considering the other calibrator SNRs (except SNRs G156.2+5.7 
and G114.3+0.3 which have reliable distance and radio luminosity values) 
we see that the luminosity decreases only a bit with respect to the diameter 
(see Fig.2). Does 
radio luminosity of SNRs actually change only a little bit until their 
diameters 
reach to values 40-50 pc? 

If, during the evolution of SNRs, L(F) value does not 
really change much then the slope of the $\Sigma$-D relation must be $-$2 
(i.e. $\Sigma$$\sim$D$^{-2}$), but no one has claimed that such an 
equation for SNRs is valid, yet. In this work, such an equation is not 
valid for all the calibrator SNRs, either. 

The most contribution for the luminosity seen to be almost constant is 
mainly due to the SNRs W44 (G34.7-0.4), W28 (G6.4-0.1) and RCW 89 
(G320.4-1.2) 
which are in very high-density medium. As it is known that, the ejected 
mass of Type I SN is less than the ejected mass of Type II SN, but 
expansion velocity of Type I SNe's remnants is greater, about 15000 km/s 
on average. If such a remnant expands in a very low-density medium (as 
often is the case), 
then in a very short time it can reach to a large diameter of about SNR 
SN1006's diameter. For such a case, dense ultra-relativistic  particles 
and high magnetic fields can not be expected. So that, such SNRs have low 
luminosity and surface brightness values corresponding to their diameters 
in the L-D and $\Sigma$-D diagrams. 

This is also true for SNR G4.5+6.8 (Kepler, SN1604) which is in the 
galactic center direction and 570 pc above the galactic plane. As 
Kepler's age and diameter are 400 yr and 4 pc, respectively, its radio 
luminosity being 5.5 times higher than SN1006's radio luminosity is 
normal (see Fig.2). This shows that for S-type SNRs expanding through 
low-density medium the luminosity decreases considerably, about 5 times, 
even only in 1000 years. 

As seen in Fig.1, the S-type SNR G93.3+6.9 (DA530, z=420 pc), which is 
the third most distant SNR from the galactic plane (after Kepler and SN 
1006), also is well below the $\Sigma$-D line.  
This Ia type SN is assumed to have been formed 5000 
years ago. Its explosion energy is 3.9$\times$10$^{50}$ erg. Its diameter 
is 26 pc (Landecker et al. 1999). Since this SNR is in a very low-density 
medium, it has a low luminosity (see Fig.2). It must be noted that,  
errors in age and explosion energy values of SNRs are not small, because 
these values are found using some not-so-precise observational data and 
theoretical models. 

The other S-type SNRs with respect to their distances from the galactic 
plane are, respectively: G166.0+4.3 (VRO 42.05.01, z=277), G296.5+10 
(PKS1209-51/52, z=261), and G119.5+10.2 (CTA 1, z=248 pc). Among these, 
G166.0+4.3 is located in the outer (Persei) arm of the Galaxy. 
There is a sharp increase in reddening values of the stars located 
between 3-4 kpc in 
the direction of this SNR (Neckel \& Klare 1980) that the density of dust 
in this region should be high. So, this SNR can not be assumed to expand 
in a low-density medium. On the other hand, since SNRs 
G296.5+10 and G119.5+10.2 are absolutely out of star formation regions, they 
can be assumed to be in lower density medium. Because of this, their 
surface brightness and luminosity values are low as expected (see 
Fig.1 and Fig.2). 

As seen in Table 1 and Figs. 1 and 2, SNR G156.2+5.7 has the lowest 
surface brightness and radio luminosity values among the calibrator 
SNRs. We can say that Kepler and SN 1006-like SNRs will have such low 
surface brightness and radio luminosity values when their diameters 
reach to about 60-70 pc. 

In $\Sigma$-D and L-D diagrams, SNRs G156.2+5.7, Kepler and SN1006 are 
roughly on the same lines that these lines can be assumed to be the 
evolutionary tracks of S-type SNRs which evolve in very low-density medium. 
The equations of these tracks are (D in pc): 
\begin{equation}
\Sigma=5 \times 10^{-17} D^{-3.32} \hspace{0.3cm} Wm^{-2}Hz^{-1}ster^{-1}
\end{equation}
\begin{equation}
L=3 \times 10^4 D^{-1.16} \hspace{0.3cm} Jy \hspace{0.1cm} pc^2
\end{equation}

From the definitions $\Sigma$ $\sim$ F/$\theta$$^2$ and L $\sim$ F d$^2$, 
the relation between $\Sigma$ and L is:
\begin{equation}
\Sigma \sim L/D^2
\end{equation}
Using the $\Sigma$-D equations (4) and (5) for the SNRs having diameters 
greater than 
and less than 36.5 pc, and Eqn.10 we can find 2 equations; one for the SNRs 
having smaller diameters:
\begin{equation}
L \sim D^{-0.47}
\end{equation}
and the other one for the SNRs having larger diameters:
\begin{equation}
L \sim D^{-4}
\end{equation}
If we compare these 2 equations with Eqns. (6) and (7), respectively, we see 
that the values of powers are very close to each other. Therefore, Eqns. 
(6) and (7) can be considered to be correct, in principle.  
   
\section{Theoretical Basis}
As known from synchrotron radiation theory, radiation (in radio band) 
per unit volume at a certain frequency (spectral density) is given as
(Ginzburg 1981)
\begin{equation}
j_\nu = 1.35\times 10^{-27} b(\alpha) (6.26\times 10^4)^\alpha K_e 
B^{\alpha+1}_{-5} \nu^{-\alpha}_{GHz} \hspace{0.2cm}erg 
\hspace{0.1cm}cm^{-3} s^{-1} ster^{-1} Hz^{-1} 
\end{equation}
Here, b($\alpha$) is a function of $\alpha$, B is the 
magnetic field of the region which emits in units of 10$^{-5}$ Gauss,  
$\nu$$_{GHz}$ is frequency of radiation in units of GHz, K$_e$ is the 
coefficient in the energy spectrum (distribution) of ultrarelativistic 
electrons.
\begin{equation}
N_e dE = K_e E^{-\gamma} dE \hspace{0.2cm}electron/cm^3
\end{equation}
Here, $\gamma$ = 2$\alpha$ + 1. In the shells of S- and C-type SNRs 
electrons are accelarated mainly by regular (Bell-Krymsky) mechanism 
(Bell 1978a, 1978b; Krymsky 1977). In order to apply this mechanism to 
SNRs, it is necessary to choose acceptable values for the Bell 
coefficients (Allakhverdiev et al. 1986d), but here we consider the 
accelaration under strong shock propagation, in general. This mechanism 
gives the number and 
energy distribution of electrons, which are accelarated in the strong 
shock wave, in units of density of the medium and velocity of the shock 
wave. K$_e$, given in Eqn.14, depends on volume density of unaccelarated 
electrons in the shock wave (n) and velocity of the shock wave (V): 
\begin{equation}
K_e \sim n V^{2\alpha}
\end{equation}
If we apply the regular mechanism to the accelaration in the SNR's shock 
wave we get:
\begin{equation}
j_\nu = 9.69\times 10^{-30} b(\alpha) (3.41\times 10^{-9})^\alpha 
\hspace{0.05cm}\Phi_e \hspace{0.05cm}
\alpha \hspace{0.05cm}n V^{2\alpha}_{8} B^{\alpha+1}_{-5} 
\nu^{-\alpha}_{GHz} 
\hspace{0.2cm}erg \hspace{0.1cm}cm^{-3} s^{-1} ster^{-1} Hz^{-1}
\end{equation}
Here, V$_8$ is the shock wave velocity in units of 10$^8$ cm/s and 
$\Phi$$_e$ is the constant given by Bell. 

As seen from Eqn.16, the radiation of the SNR shell's unit volume at a 
certain frequency (j$_{\nu}$) is related to n, V, and B for radio 
radiation with spectral index, $\alpha$=0.5 (note that the spectral 
indices of S and C-type SNRs are always close to 0.5 and this confirms 
the Bell mechanism): 
\begin{equation}
j_{\nu} \sim n\hspace{0.1cm}VB^{1.5}
\end{equation}

Using the observational data of SNRs in X-ray, optical, and radio bands 
Tagieva (2002) showed that
\begin{equation}
n \sim D^{-0.9\pm 0.4}, \hspace{0.3cm} V \sim D^{-1.3\pm 0.3}, 
\hspace{0.3cm} B \sim D^{-0.8}
\end{equation}
Since the magnetic field values are known (with large errors) only 
of a few of the SNRs, it is not yet possible to find a relation between B 
and D directly from observations. If the dynamo mechanism to increase the 
magnetic field is not working, then, as magnetic field is freezed to gas, 
B$\sim$n$^{2/3}$$\sim$D$^{-0.6}$. If 
the dynamo is also working a bit, we can roughly assume that 
B$\sim$D$^{-0.8}$ as in Eqn.18. Using the relations of Eqn.18 in Eqn.17:
\begin{equation}
j_{\nu} \sim D^{-3.4}
\end{equation}

Shock wave front expands like a shell that its volume increases roughly 
with D$^2$. Because of this, luminosity is:
\begin{equation}
L_{\nu} \sim D^2 j_{\nu}
\end{equation}
From Eqns. 19 and 20:
\begin{equation}
L \sim D^{-1.4}
\end{equation}
This is roughly in agreement with the average power of the L-D 
equations (Eqns. 6 and 7) found from the L-D diagram (Fig.2). But using 
this single equation, instead of using the 2 L-D equations, leads to 
results with larger errors for some of the SNRs.

As known from observations, while SNR's diameter is increasing most of the 
parts of the shock wave front do not interact with interstellar clouds, so 
that, behind the wave front some large low-density regions form. In these 
regions, as the value of magnetic field intensity is smaller, high-energy 
ultrarelativistic electrons can hardly be trapped. 
The electrons in denser regions of the wave front 
move along the magnetic field lines, which are very disordered, and, after 
reaching the low-density regions, they can also leave the SNR. 
It must be noted that, for SNRs with D$\sim$40 pc the magnetic field, on 
average, is (2-6) $\mu$G (Seward et al. 1995; du Plessis et al. 
1995; for G93.7-0.3 B=2.3 $\mu$G, Uyan\i ker et al. 2001). 
The lifetime of ultrarelativistic electrons is:
\begin{equation}
t(yr) \cong \frac{3 \times 10^2}{H^2(Gauss) E(eV)}
\end{equation}
Energy and lifetime of the ultrarelativistic electrons, which radiate in 
such magnetic fields at 1 
GHz, are not less than $\sim$5$\times$10$^3$ MeV and 4$\times$10$^4$ 
yr; the lifetime of ultrarelativistic electrons is comparable to the 
average lifetime of SNRs that, not only the new-accelarated electrons 
but mainly the electrons which have already been accelarated do produce 
the radiation of SNRs. As these electrons leave the low-magnetic field 
regions of the SNR, the SNR's radiation decreases rapidly. So that, 
in most of the cases, the part of the shell, which is farther away from 
the Galactic plane (where the number density of clouds is low), is less 
bright (Caswell \& Lerche 1979; Allakhverdiev et al. 1983b). Since such a 
decrease in the radiation is more effective in large-diameter SNRs, the 
slope of the $\Sigma$-D equation for such SNRs should be larger.  

\section{Discussion and Conclusions}
In this work, we have constructed the $\Sigma$-D relation for C and 
S-type SNRs. We have also adopted distance and diameter values for all of 
the Galactic SNRs given in the Galactic SNRs catalog of Green (2001) and 
also for some recently found SNRs. Although the SNR nature of Sgr A East 
(G0.0+0.0) and G10.0-0.3 is uncertain, we have included them in Table 1, 
because they are given in the Galactic SNRs catalog (Green 2001). 

As mentioned in the introduction, in the last 45 years many $\Sigma$-D 
relations were constructed and presented in the literature. In some of 
these works a single linear dependence between Log $\Sigma$ and Log D 
values of calibrator SNRs was given. In other works 2 linear dependences 
(with 2 different slopes) were given. The most recent $\Sigma$-D dependence 
was given by Case \& Bhattacharya (1998): $\Sigma$(1 GHz) = 
2.07$^{+3.10}_{-1.24}$ $\times$ 10$^{-17}$ D$^{-2.38\pm 0.26}$ 
Wm$^{-2}$Hz$^{-1}$ster$^{-1}$ (Cas A is not included). 

As seen above, in Case \& Bhattacharya (1998) the relation between $\Sigma$
and D is given with only one equation, instead of two, for the whole set 
of calibrators. The reason of this is that, for some of the calibrator 
SNRs having small surface brightness values ($\Sigma$ $<$ 10$^{-21}$ 
Wm$^{-2}$Hz$^{-1}$ster$^{-1}$), they assume 
diameter (distance) values larger than the values which we have adopted 
to construct our $\Sigma$-D relation (which includes 2 equations with 2 
different slopes). So, for small-$\Sigma$ calibrator SNRs the adopted 
diameters (distances) given in Case \& Bhattacharya (1998) are larger 
than the diameters (distances) which we have adopted. It is necessary to 
compare and discuss the SNRs for which the differences in the adopted 
distance values are the largest. For example, for SNRs G156.2+5.7, 
G166.0+4.3, G180.0-1.7, and G205.5+0.5 the distances adopted by Case \& 
Bhattacharya (1998) and by us (given in brackets) are, respectively: 3.0 
(2.0) kpc, 4.5 (3.8) kpc, 1.5 (1.0) kpc, and 1.6 (1.0) kpc. In Section 
2.3, we discussed the distance values of these calibrator SNRs in detail.  

There are 3 SNRs which were considered as calibrators 
by Case \& Bhattacharya (1998), but excluded by us in constructing the 
$\Sigma$-D dependence: G116.5+1.1, G160.9+2.6, and G166.2+2.5. The 
adopted diameter values of these SNRs given in Case \& Bhattacharya 
(1998) differ significantly from our adopted diameter values (Table 1).
The distance values of these SNRs according to Case 
\& Bhattacharya (1998) and the distance values of the SNRs adopted by us
(given in brackets) are, respectively: 5 (3.5) kpc, 3.0 (1.2) kpc, and 
4.5 (2.7) kpc. We will discuss the distance values of these 3 SNRs below.

For SNR G116.5+1.1 d=3.6-5.2 kpc (Green 2001) and d=4.4 kpc (Reich \& 
Braunsfurth 1981; Lorimer et al. 1998) values were given. In the 
direction of 
this SNR, there is Persei arm of the Galaxy about 3-4 kpc distant from 
the Sun. On the other hand, there is no star formation region 
at $\sim$5 kpc in this direction. In this part of the interstellar 
medium which is not dense (Fesen et al. 1997), this SNR might reach a 
diameter of about 70-90 pc at d=3.5-4.4 kpc.  

The supercavity in which SNR G116.5+1.1 is located also include 
G114.3+0.3 and G116.9+0.2 (Fich 1986). The diameter (distance) of SNR 
G116.5+1.1 is not expected to be larger than the diameter (distance) of SNR 
G114.3+0.3, because the surface brightness of SNR G116.5+1.1 is greater 
than the surface brightness of SNR G114.3+0.3. Also, a few SNRs being 
located in the same region requires them to be in the Galactic arm. From 
our $\Sigma$-D relation the distance of this SNR is found to be 2.7 kpc. 
Taking this and the distance values given above into account we have 
adopted d=3.5 kpc for SNR G116.5+1.1. 

For SNR G160.9+2.6 d=1.7 kpc (Braun et al. 1989) and d$<$4 kpc (Green 
2001) were given. From the $\Sigma$-D relation d=1.2 kpc and this value 
is adopted as the distance of SNR G160.9+2.6. In this direction, between 
d=1-3 kpc, the interstellar absorption is almost constant (Neckel \& 
Klare 1980). This shows that the medium (d=1-3 kpc) has a very low 
density. In this part of the Galaxy there is no star formation region. 
So, even for small diameter values, the surface brightness of this SNR 
must be small. 

For SNR G166.2+2.5 d=8 kpc (Routledge et al. 1986; Green 2001), d=4.5 kpc 
(Landecker et al. 1989), and d=2 kpc (Braun et al. 1989) were given. 
Since, there is no star formation region at these distances in this 
direction, this SNR can not be much above the $\Sigma$-D line. A distance 
of 2.5 kpc is found from the $\Sigma$-D relation and this value is 
adopted as the distance of this SNR. 

Above, we discussed the escape of relativistic electrons from 
large-diameter SNRs. Because of this effect, the SNR's luminosity, and 
also its surface brightness, should decrease rapidly with 
respect to the diameter of the SNR. So, the $\Sigma$-D dependence of 
large-diameter SNRs should be sharper (with a larger slope) compared to 
the the $\Sigma$-D dependence of the SNRs which have smaller diameters. 

There is also another such effect; at the initial stages of the evolution, 
SNRs expand within the HII regions which were created by the progenitors of 
the SNRs. During this time, as the shock wave has a high velocity its 
temperature is also high that the X-ray radiation will be high. Expansion 
velocity of the SNR must decrease a lot when the SNR's shock wave reachs 
the HII region's boundary and interacts with a dense neutral gas and 
molecular clouds (Lozinskaya 1981; Chevalier 1999; Eikenberry 2002). 
But at the same time, the rate of drop of the X-ray radiation should 
decrease as the mass and density of the gas in the shock wave increases. 
Naturally, a SNR which has a high explosion energy and which is expanding 
through a dense medium must have a higher X-ray luminosity compared to a 
SNR with a lower explosion energy expanding through a lower-density 
medium. It is seen that, after the SNR reachs the boundary of the HII 
region, which was formed by the progenitor O-type star, a small increase 
in the SNR's size will be accompanied by a sharp decrease in the SNR's 
radio and X-ray luminosity. Depending on sizes of the HII regions, 
surface brightness of SNRs will begin to drop sharply at different values 
of the diameter. We have assumed a value of D=36.5 pc as the turn-off 
point in the $\Sigma$-D relation (which is not an evolutionary track), i.e. 
after the SNR reachs a diameter of roughly about 36.5 pc the slope of 
the $\Sigma$-D relation sharply changes.

\clearpage
{\bf References} \\
Ahumada, J. and Lapasset, E. 1995, A\&AS, 109, 375 \\
Albinson, J.S., Tuffs, R.J., Swinbank, E., Gull, S.F. 1986, MNRAS, 219, 
427 \\
Allakhverdiev, A.O., Amnuel, P.R., Guseinov, O.H., Kasumov, F.K. 1983a,
Ap\&SS, 97, 261 \\
Allakhverdiev, A.O., Amnuel, P.R., Guseinov, O.H., Kasumov, F.K. 1983b,
Ap\&SS, 97, 287 \\
Allakhverdiev, A.O., Guseinov, O.H., Kasumov, F.K. 1986a, Astrofizika,
24, 97 \\
Allakhverdiev, A.O., Guseinov, O.H., Kasumov, F.K., \& Yusifov, I.M. 
1986b, Ap\&SS, 121, 21 \\
Allakhverdiev, A.O., Guseinov, O.H., Kasumov, F.K. 1986c, Astrofizika, 
24, 397 \\
Allakhverdiev, A.O., Asvarov, A. I., Guseinov, O.H., Kasumov, F.K. 1986d,
Ap\&SS, 123, 237 \\
Allakhverdiev, A. O., Alpar, M. A., G\"{o}k, F., Guseinov, O.H. 1997, Tur. 
Jour. of Phys., 21, 688 \\
Allen, G.E., Petre, R., Gotthelf, E.V. 1998, AAS 193, 5101 \\
Ankay, A. \& Guseinov, O.H. 1998, A\&ATr, 17, 301 \\
Asaoka, I. \& Aschenbach, B. 1994, A\&A, 284, 573 \\
Asvarov A.I., Dogiel V.A., Guseinov O.H., Kasumov F.K. 1990, A\&A, 229, 
196 \\
Ayd\i n, C., Albayrak, B., Ankay, A., Guseinov, O.H. 1997, Tr. J. of 
Physics, 21, 857 \\
Bamba, A., Koyama, K., Tomida, H. 2000, PASJ, 52, 1157 \\
Bandiera, R. 1987, ApJ, 319, 885 \\
Bandiera, R., Pacini, F., \& Salvati, M. 1996, ApJ, 465, L39 \\
Becker W., Brazier K., Trumper J. 1996, A\&A, 306, 464 \\
Becker, R.H., Helfand, D.J., Szymkowiak, A.E. 1982, ApJ, 255, 557 \\
Bell, A. R. 1978a, MNRAS, 182, 147 \\
Bell, A. R. 1978b, MNRAS, 182, 443 \\
Berdnikov, L. N. \& Efremov, Y. N. 1993, Pisma Astronomicheskii Zhurnal, 
19, 957 \\
Berkhuijsen, E.M. 1986, A\&A, 166, 257 \\
Berthiaume, G.D., Burrows, D.N., Garmire, G.P., \& Nousek, J.A. 
1994, ApJ 425, 132 \\
Blaha \& Humphreys 1989, AJ, 98, 1598 \\
Blair, W.P., Sankrit, R., Raymond, J.C. and Long, K.S. 1999, AJ, 118, 942 
\\ 
Blair, W. B., Raymond, J. C., Kriss, G. A. 1995, ApJ, 454, L53 \\
Blandford, R.D. \& Cowie, L.L. 1982, ApJ, 260, 625 \\
Bocchino, F., Maggio, A., Sciortino, S. 1999, A\&A, 342, 839 \\
Borkowski, K.J., Blondin, J.M., Sarazin, C.L. 1992, ApJ, 400, 222 \\
Borkowski, K.J., Rho, J., Reynolds, S.P., Dyer, K.K. 2001, ApJ, 550, 334 
\\ 
Braun, R. 1987, A\&A, 171, 233 \\
Braun, R. \& Strom, R.G. 1986, A\&A, 164, 208 \\
Braun, R., Goss, W.M., Lyne, A.G. 1989, ApJ, 340, 355 \\
Brazier, K.T.S. \& Johnston, S. 1999, MNRAS, 305, 671 \\
Brazier, K.T.S., Kanbach, G., Carraminana, A. 1996, MNRAS, 281, 1033 \\
Brazier, K.T.S., Reimer, O., Kanbach, G., Carraminana, A. 1998, MNRAS, 295,
819 \\
Camilo, F., Stairs, I. H., Lorimer, D. R., et al. 2002, astro-ph/0204219 
\\ 
Carter,L. M., Dickel, J. R., \& Bomans, D. J. 1997, PASP 109, 990 \\
Case G.L. \& Bhattacharya, D. 1998, ApJ, 504, 761 \\
Caswell J.L., Kesteven M.J., Komesaroff M.M., et al. 1987, MNRAS, 225, 329 
\\ 
Caswell, J.L., Murray, J.D., Roger, R.S., et al. 1975, A\&A, 45, 239 \\
Caswell, J.L., \& Lerche, I. 1979, MNRAS, 187, 201 \\
Cha, A.N., Sembach, K.R., Danks, A.C. 1999, ApJ, 515, L25 \\
Chakrabarty, D., Pivovaroff, M.J., Hernquist, L.E. et al. 2001, ApJ, 548, 
800 \\
Chevalier, R.A., Krishner, R.P., Raymond, J.C. 1980, ApJ, 235, 186 \\ 
Chevalier, R.A. 1999, ApJ, 511, 798 \\
Clark, D.H. \& Caswell, J.L. 1976, MNRAS, 174, 267 \\
Claussen M.J., Frail D.A., Goss W.M., Gaume R.A. 1997, ApJ, 489, 143 \\
Claussen M.J., Goss W.M., Frail D.A., \& Seta, M. 1999a, AJ, 117, 1387 
\\ 
Claussen M.J., Goss W.M., Frail D.A., Desai K. 1999b, ApJ, 522, 349 \\
Cox, D.P., Shelton, R.L., Maciejewski, W., et al. 1999, ApJ, 524, 179 \\ 
Craig, W.W., Hailey, C.J., Pisarski, R.L. 1997, ApJ, 488, 307 \\
Danks, A. C. 2000, Ap\&SS, 272, 127 \\
Davelaar, J., Smith, A., Becker, R. 1986, ApJ, 300, L59 \\
De Jager, O. C., Harding, A. K., Strickman, M. S. 1996, ApJ, 460, 729 \\
Denoyer, L. K. 1979a, ApJ, 228, L41 \\
Denoyer, L. K. 1979b, ApJ, 232, L165 \\
Denoyer L.K. 1983, ApJ, 264, 141 \\
Du Plessis, I., de Jager, D.C., Buchner, S., et al. 1995, AJ, 453, 746 \\ 
Dickel J. R., Green, A., Ye, T., \& Milne, D. K. 1996, AJ, 111, 340 \\
Dickel, J.R., Strom, R.G., Milne, D.K. 2001, ApJ, 546, 447 \\
Diplas, A. \& Savage, B.D. 1994, ApJS, 93, 211 \\
Dubner, G.M., Velazquez, P.F., Goss, W.M., Holdaway, M.A. 2000, AJ, 120, 
1933 \\
Efremov, Y. N. 1989, Ochagi zvezdoobrazovaniya v galaktikakh (Sites of 
star formation in galaxies), Moscow, Nauka. \\
Eikenberry, S.S. 2002, Invited review for the Woods Hole 2001 Conf. on 
GRBs/SGRs (astro-ph/0203054) \\
Favata, F., Vink, J., Dal Fiume, D., et al. 1997, A\&A, 342, L49 \\ 
Fesen, R.A. 1984, ApJ, 281, 658 \\
Fesen, R.A., Blair, W.P., Kirshner, R.P., et al. 1981, ApJ, 247, 148 \\
Fesen, R.A. \& Hurford, A.P. 1995, AJ, 110, 747 \\
Fesen, R.A., Winkler, P.F., Rathore, Y., et al. 1997, AJ, 113, 767 \\ 
Fesen, R.A., Wu, C.-C., Leventhal, M., Hamilton, A.J.S. 1988, ApJ, 327, 
164 \\
Fesen, R.A., Downes, R.A. and Wallace, D. 1995, AJ, 110, 2876 \\
Fich M. 1986, ApJ, 303, 465 \\
Fink, H. H., Asaoka, I., Brinkmann, W., Kawai, N., \& Koyama, K. 1994, 
A\&A, 283, 635  \\
Frail D.A., Goss W.M., Reynoso E.M., et al. 1996, AJ,  111, 1651 \\
Frail D.A., Goss W.M., Slysh V.I. 1994a, ApJ, 424, L111 \\
Frail, D.A., Kassim, N.E., Weiler, K.W. 1994b, AJ, 107, 1120 \\
Frail D.A., Kassim N.E., Cornwell T.J. \& Goss W.M. 1995, ApJ  454, L129 
\\ 
Frail D.A. \& Kulkarni S.R. 1991, Nature, 352, 785 \\
Frail, D.A. \& Moffett, D.A. 1993, ApJ, 408, 637 \\
Fujimoto, R., Tanaka, Y., Inoue, H., et al. 1995, PASJ, 47, L31 \\
Furst, E., Reich, W., Seiradakis, J. H. 1993, A\&A, 276, 470 \\
Gaensler, B. M., Gotthelf, E. V., \& Vasisht, G. 1999a, ApJ, 526, L37 \\ 
Gaensler, B.M., Brazier, K.T.S., Manchester, R.N. et al. 1999b, MNRAS, 
305, 724 \\
Gaensler, B.M. 2000, Pulsar Astronomy-2000 and Beyond, ASP Conference
Series, vol.202, p.703 \\
Gaensler, B.M., Dickel, J.R., Green, A.J. 2000a, ApJ, 542, 380 \\
Gaensler, B. M., Bock, D. C. -J., Stappers, B. W. 2000b, ApJ, 537, L35 \\
Galas, C.M.F., Tuohy, I.R., Garmire, G.P. 1980, ApJ, 236, L13 \\
Galas, C.M., Tuohy, I.R., Garmire, G.P. 1980, ApJ, 119, 281 \\
Garmany, C.D. \& Stencel, R.E. 1992, A\&AS, 94, 211 \\
Giacani, E. B., Dubner, G. M., Kassim, N. E., et al. 1997, AJ, 113, 1379 
\\ 
Giacani, E.B., Dubner, G.M., Green, A.J., Goss, W.M., \& Gaensler, B.M.
2000, AJ, 119, 281 \\
Ginzburg, V. L. 1981, Theoreticheskaya fizika i astrofizika, Izd. Nauka, 
Moscow  \\ 
Gotthelf, E. V. \& Vasisht, G. 1998, NewA, 3, 293 \\
Gotthelf, E. V., Vasisht, G., \& Dotani, T. 1999a, 522, L49 \\
Gotthelf, E.V, Petre, R., Vasisht, G. 1999b, ApJ, 514, L107 \\
Gotthelf, E.V., Petre, R., Hwang, U. 1997, ApJ., 487, L175 \\
Gray, A. D., Landecker, T. L., Dewdney, P. E., Taylor, A. R. 1999, ApJ, 
514, 221 \\ 
Green, D.A. 1984, MNRAS, 209, 449 \\
Green, D.A. 1989, MNRAS, 238, 737 \\
Green, D.A. 2000. 'A Catalogue of Galactic Supernova Remnants
(2000 August version) \\
Green, D.A., 2001. 'A Catalogue of Galactic Supernova Remnants (2001
December version), http://www.mrao.cam.ac.uk/surveys/snrs/ \\
Greidanus, H. \& Strom, R.G. 1992, A\&A, 257, 265 \\
Greiveldinger C, Caucino S,  Massaglia S., et al. 1995, ApJ, 454, 855 \\ 
Guo, Z. \& Burrows, D.N. 1997, ApJ, 480, L51 \\
Guseinov, O. H., Yerli, S. K., \"{O}zkan, S., Sezer, A., Tagieva, S. O. 
2002, preprint \\ 
Hailey, C.J. \& Craig, W.W. 1994, ApJ, 434, 635 \\
Hamilton, A. J., Fesen, R. A., Wu, C. C., et al. 1997, ApJ, 482, 838 \\
Helfand, D.J., Becker, R.H., White, R.L. 1995, ApJ, 453, 741 \\
Huang, Y.-L. \& Thaddeus, P. 1985, ApJ, 295, L13 \\
Huang, Y.-L. \& Thaddeus, P. 1986, ApJ, 309, 804 \\
Hughes, J.P., Helfand, D.J., Kahn, S.M. 1984, ApJ, 281, L25 \\
Hulleman, F., van Kerkwijk, M. H., Verbunt, F. W. M., \& Kulkarni, S. R.
2000, A\&A, 358, 605 \\
Humphreys, R.M. 1978, A\&AS, 38, 309 \\
Hwang, U., Petre, R., Hughes, J.P. 1999, AAS, 194, 8513 \\ 
Israel, F.P. 1980, A\&A, 90, 246 \\
Junkes, N., Furst, E., Reich, W. 1992, A\&AS, 96, 1 \\
Kahn, S. M., Gorenstein, P., Harnden, F. R., Jr., Seward, F. D. 1985, 
ApJ, 299, 821 \\
Kaplan, D.L., Kulkarni, S.R., Murray, S.S. 2001, arXiv:astro-ph/0102054v3 
\\ 
Kaspi, V. M., Chakrabarty, D., \& Steinberger, J. 1999, ApJ 525, L33 \\
Kaspi, V.M., Manchester, R.N., Johnston, S., et al. 1996, AJ, 111, 2028 
\\ 
Kaspi, V.M., Lyne, A. G., Manchester, R. N. et al. 1993, ApJ, 409, L57  \\
Kaspi, V.M. 2000, Pulsar Astronomy-2000 and beyond. ASP Conference Series- 
Eds. Kramer M, Wax, N. and Wielebinski, N., p.485 \\
Kellett, B.J., Branduardi-Raymont, G., Gulhane, J.L., et al. 1987, MNRAS,
225, 199 \\
Keohane, J., Petre, R., Gotthelf, E. V., et al. 1997, ApJ, 484, 350 \\
Kirshner R.P., Winkler P.F. \& Chevalier R. A. 1987, ApJ, 315,
L135 \\
Koralesky, B., Frail, D.A., Goss, W.M., Claussen M.J., \& Green, A.J.
1998a, AJ, 116, 1323 \\
Koralesky, B., Rudnick, L., Gotthelf, E. V., \& Keohane, J. W. 1998b, 
ApJ, 505, L27 \\
Koyama K., Petre R., Gotthelf E.V., et al. 1995, Nature, 378, 255 \\
Krymsky, G. F. 1977, Dokl. Akad. Nauk SSSR, 234, 1306 \\
Laming, J. M., Raymond, J. C., McLaughlin, B. M., Blair, W. P. 1996, 
ApJ, 472, 267 \\
Landecker, T.L., Pineault, S., Routledge, D., Vaneldik, J.F. 1989, MNRAS,
237, 277 \\
Landecker, T.L., Roger, R.S., Higgs, L.A. 1980, A\&AS, 39, 133 \\
Landecker, T.L., Routledge, D., Reynolds, S.P., et al. 1999, ApJ, 527, 
866 \\
Li, Z.W., \& Wheeler, J.C. 1984, BAAS, 16, 334 \\
Long K.S., Blair W.P., White R.L., Matsui Y. 1991, ApJ, 373, 567 \\
Long, K.S., Blair, W.P., van den Bergh, S. 1988, ApJ, 333, 749 \\
Long, K. S. \& Blair, W. P. 1990, ApJ, 358, L13 \\
Lorimer, D.R., Lyne, A.G. \& Camilo, F. 1998, A\&A, 331, 1002 \\
Lozinskaya, T.A. 1981, Soviet Astron. Lett., 7, 17 \\
Lozinskaya, T.A., Pravdikova,V.V., Finoguenov, A.V., et al. 2000, AstL,
26, 77 \\
Lyne, A. G., Pritchard, R. S., Graham-Smith, F., Camilo, F. 1996, Nature, 
381, 497 \\
Lynga, G. 1987, Catalogue of open clusters (5. edition), Lund Observatory 
\\ 
Marsden, D., Lingenfelter, R., Rothschild, R., \& Higdon, J. 1999, AAS, 
195, 26.05 (astro-ph/9912315) \\
Mavromatakis, F., Papamastorakis, J., Paleologou, E.V., \& Ventura, J.
2000, A\&A, 353, 371 \\ 
Meaburn, J. \& Allan, P.M. 1986, MNRAS, 222, 593 \\
Melnik, A.M. \& Efremov, Yu. N. 1995, AstL, 21, 10 \\
Mereghetti, S., Bignami, G.F., Caraveo, P.A. 1996, ApJ, 464, 842 \\
Mills, B.Y., Turtle, A.J., Little, A.G., Durdin, J.M. 1984, Austral. J.
Phys., 37, 321 \\
Milne, D.K. 1979, Australian J. Phys., 32, 83 \\
Minkowski, R. 1958, Rev. Mod. Phys., 30, 1048 \\
Morini, M., Robba, N. R., Smith, A., Van der  klis, M. 1988, ApJ, 333,
777 \\
Murray, S. S., Ransom, S. M., Juda, M. et al. 2002, ApJ, 566, 1039 \\
Neckel, Th. \& Klare, G. 1980, A\&AS, 42, 251 \\
Nugent, J., Pravdo, S., Garmire, G., et al. 1984, ApJ, 284, 612 \\
Odegard, N. 1986, ApJ, 301, 813 \\
\"{O}gelman, H. \& Buccheri, R. 1987, A\&A, 186, L17 \\
\"{O}gelman, H., Koch-Miramond, L., Auriere, M. 1989, ApJ, 342, L83 \\ 
Olbert, C.M., Clearfield, C.R., Williams, N.E., et al. 2001, ApJ, 554, 
L205 \\
Oliva, E., Moorwood, A.F.M., Danziger I.J. 1990, A\&A, 240, 453 \\
Oliva, E., Moorwood, A.F.M., Drapatz, S., et al. 1999, A\&A, 343, 943 \\
Parmar, A.N., Oosterbroek, T., Favata, F., et al. 1998, A\&A 330, 175 \\ 
Patel, S.K., Kouveliotou, C., Woods, P.M., et al. 2001, ApJ, 563, L45 \\
Pavlov, G.G., Zavlin, V.E., Trumper, J. 1999, ApJ, 511, L45 \\
Petre, R., Becker, C.M., Winkler, P.F. 1996, ApJ, 465, L43 \\
Petruk, O. 1999, A\&A, 346, 961 \\
Pfeffermann, E., Aschenbach, B., Predehl, P. 1991, A\&A, 246, L28 \\
Phillips, A.P., Gondhalekar, P.M., \& Blades, J.C. 1981, MNRAS, 195, 485 
\\ 
Pineault, S., Landecker, T.L., Madore, B., Gaumont-Guay, S. 1993, AJ, 105,
1060 \\
Pineault, S., Landecker, T.L., Routledge, D. 1987, ApJ, 315, 580 \\   
Poveda, A. \& Woltjer, L. 1968, AJ, 73, 65 \\
Predehl, P. \& Trumper, J. 1994, A\&A, 290, L29 \\   
Radhakrishnan V., Goss W.M., Murray J.D., Brooks J.W. 1972, ApJS, 24, 49 
\\ 
Raymond, J. C., Blair, W. P., Long, K. S., et al. 1997, ApJ, 482, 881 \\
Reach, W.T. \& Rho, J. 1998, ApJ, 507, L93 \\
Reed, J.E., Hester, J.J., Fabian, A.C., Winkler, P.F. 1995, ApJ, 440, 706 
\\ 
Reich, W. \& Braunsfurth, E. 1981, A\&A, 99, 17 \\
Reich, W., Furst E., Arnal, E.M. 1992, A\&A, 256, 214 \\
Reynolds S. P., Lyutikov, M., Blandford, R. D., Seward, F. D. 1994, 
MNRAS, 271, L1 \\ 
Reynolds, S.P. 1996, ApJ, 459, L13 \\
Reynoso E. M. , Dubner G.M., Goss W.M., Arnal E.M. 1995, AJ, 110, 318 \\
Reynoso, E.M., Moffett, D.A., Goss, W.M., et al. 1997, ApJ, 491, 816 \\
Reynoso, E.M., Velazquez, P.F., Dubner, G.M., Goss, W.M. 1999, AJ, 117,  
1827 \\
Reynoso, E.M. \& Goss, W.M. 1999, AJ, 118, 926 \\
Rho, J. \& Petre, R. 1996, ApJ, 467, 698 \\
Rho, J. \& Petre, R. 1997, ApJ, 484, 828 \\
Rho, J. \& Petre, R. 1998, ApJ, 503, L167 \\
Rho, J. H., Petre, R., Schlegel, E. M., \& Hester, J. 1994, ApJ, 430, 757 
\\ 
Rho, J., Decourchelle, H., Petre, R. 1998, AAS, 193, 7401 \\
Rho, J.H. \& Petre, R. 1993, American Astronomical Society Meeting 183, 
101.07 \\
Rho, J.-H., Petre, R., Pisarski, R., Jones, L.R. 1996, Max-Planck Institut
fur Extraterres. Trische Physik report 1996, 263, 273 \\
Rho, J., Jarrett, T.H., Cutri, R.M., Reach, W.T. 2001, ApJ, 547, 885 \\    
Roberts, D.A., Goss, W.M., Kalberla, P.M.W. et al. 1993, A\&A, 274, 427 
\\ 
Roberts, M.S.E., Kaspi, V.M., Vasisht, G. et al. 2000, AA Society,
HEAD, 32, 4411 \\
Roberts, M.S.E., Romani, R.W., Kawaii, N. et al. 2001, in The nature of 
unidentified galactic high energy gamma-ray sources, Edi. by A. 
Carraminana, O. Reimer, and D. J. Thompson (Kluwer Academic Publishers 
Dordrecht), vol.267, p.135 \\
Roger R.S. , Milne D.K., Kesteven M.J., Wellington K.J. \& Haynes R.F. 1988,
ApJ., 332, 940 \\
Rosado, M., Ambrocio-Cruz, P., Le Coarer, E., Marcelin, M. 1996, A\&A, 315, 
243 \\
Routledge, D., Landecker, T. L., Vaneldik, J. F. 1986, A\&A, 221, 809 \\ 
Routledge, D., Dewdney, P.E., Landecker, T.L., \& Vaneldik, J.F. 1991, 
A\&A, 247, 529 \\
Rowell, G.P., Naito, T., Dazeley, S.A. 2000,  A\&A, 359, 337 \\
Ruiz, M.T. 1983, AJ, 88, 1210 \\
Safi-Harb, S., \"{O}gelman, H., Finley, J. 1995, ApJ, 439, 722 \\
Sahibov, F.H. \& Smirnov, M.A. 1982, Soviet Astron. Lett., 8, 150 \\
Sahibov, F.H. \& Smirnov, M.A. 1983, AZh, 60, 676 \\
Schaefer, B.E. 1996, ApJ, 459, 438 \\
Schwarz, U.J., Goss, W.M., Kalberla, P.M., Benaglia, P. 1995, A\&A, 299, 
193 \\
Seward F.D., Harnden F.R., Szymkowiak A., Swank J. 1984, ApJ, 281, 650 \\ 
Seward, F.D., Dame, T.M., Fesen, R.A., Aschenbach, B. 1995, ApJ, 449, 
681 \\
Shklovsky, I.S. 1960, Soviet Astron., 4, 243 \\
Slane, P., Seward, F.D., Bandiera, R., et al. 1997, ApJ, 485, 221 \\
Sollerman, J., Lundqvist, P., Lindler, D., et al. 2000, ApJ, 537, 861 \\
Tagieva, S.O. 2002, accepted for publication by Astronomy Letters \\ 
Tamura, K., Kawai, N., Yoshida, A., Brinkmann, W. 1996, PASJ, 48, L33 \\
Tatematsu, K., Fukui, Y., Landecker, T.L., Roger, R.S. 1990, A\&A, 237, 
189 \\
Taylor, J.N., Manchester, R.N., Lyne, A.G., Camilo, F. 1996, A catalog of
706 PSRs \\
Torii, K., Kinugasa, K., Katayama, K., et al. 1998, ApJ, 503, 843 \\
Torii, K., Kinugasa, K., Toneri, T., et al. 1998, ApJ, 494, L207 \\
Torii, K., Tsunemi, H., Dotani, T., et al. 1999, ApJ, 523, L69 \\
Torii, K., Slane, P.O., Kinugasa, K., et al. 2000, PASJ, 52, 875 \\
Trussoni, E., Massaglia, S., Caucino, S., et al. 1996, A\&A, 306, 581 \\ 
Tuohy, I.R. \& Garmire, G.P. 1980, ApJ, 239, L107 \\
Uyan\i ker, B., Kothes, R., Brunt, C.M., 2001, astro-ph/0110001 \\
Vasisht, G., Aoki, T., Dotani, T., Kulkarni, S. R., Nagase, F. 1996, 
ApJ, 456, L59 \\
Vasisht, G. \& Gotthelf, E.V. 1997, ApJ, 486, L129 \\
Vasisht, G., Kulkarni, S.K., Anderson, S. B., et al. 1997, ApJ, 476, L43 
\\ 
Vink, J., Bloemen, H., Kaastra, J.S., Bleeker, J.A.M. 1998, A\&A, 339,
201 \\
Weiler, K.W. \& Panagia N. 1978, A\&A, 70, 419 \\ 
Willingale, R., West, R.G., Pye J.P, Stewart G.C. 1996, MNRAS, 278,
749 \\
Winkler, P.F., Canizares, C.R., Clark, G.W., et al. 1981a, ApJ, 245, 574 
\\ 
Winkler, P.F., Canizares, C.R., Clark, G.W., et al. 1981b, ApJ, 246, L27 
\\   
Winkler, P.F. \& Long, K.S. 1997, ApJ, 491, 829 \\
Woermann, B., Gaylard, M.J., Otrupcek, R., et al. 2000, MNRAS, 317, 421 
\\ 
Wooten, A. 1981, ApJ, 245, 105 \\
Wright, M., Dickel, J., Koralesky, B., Rudnick, L. 1999, AJ, 518, 284 \\
Wu, C.-C., Crenshaw, D.M., Fesen, R.A., et al. 1993, ApJ, 416, 247 \\
Yamauchi, S., Ueno, S., Koyama, K., et al. 1993, PASJ, 45, 795 \\
Yamauchi, S., Koyama, K., Tomida, H., et al. 1999, PASJ, 51, 13 \\  
Zavlin, V. E., Trumper, J., Pavlov, G. G. 1999, ApJ, 525, 959 \\
Zavlin, V. E., Pavlov, G. G., Sanwal, D., Trumper, J. 2000, ApJ, 540, L25 
\\

\begin{flushleft}
\begin{tabular}{ccccccccc}
\multicolumn{9}{l}{\bf Table 1: Data of Galactic SNRs} \\ \hline
l,b & Name & Type & $\Sigma$ & d$_{\Sigma - D}$ & D$_{\Sigma - D}$ &
d$_{ado}$ & D$_{ado}$ & L \\
& & & & (kpc) & (pc) & (kpc) & (pc) & erg/s \\
\hline
0.0+0.0 & SgrAE & S & 1.720E-18 & 3.61 & 3.05 & 8.5 & 7.2 & 3.12E+35 \\
0.3+0.0 & & S & 2.759E-20 & 5.10        & 16.25 & 5.4 & 17.2 & 2.77E+34 \\
0.9+0.1 & & C(PWN) & 4.233E-20 & 5.90 & 13.66 & 8 & 18.5 & 4.98E+34 \\
1.0-0.1 & & S & 3.527E-20 & 6.35 & 14.71 & 6.4 & 14.8 & 2.66E+34 \\
1.4-0.1 & & S & 3.010E-21 & 12.97 & 37.72 & 13 & 37.8 & 1.46E+34 \\
1.9+0.3 & & S & 6.271E-20 & 33.75 & 11.65 & 20 & 6.9 & 1.04E+34 \\
3.7-0.2 & & S & 2.248E-21 & 10.99 & 39.6 & 11 & 39.6 & 1.20E+34 \\
3.8+0.3 & & S? & 1.858E-21 & 7.81 & 40.88 & 7.8 & 40.8 & 1.05E+34 \\
4.2-3.5 & & S & 6.143E-22 & 6.04 & 49.18 & 6 & 48.8 & 4.98E+33 \\
4.5+6.8* & Kepler & S & 3.177E-19 & 7.14 & 6.04 & 4.8 & 4.1 & 1.89E+34 \\
4.8+6.2 & & S & 1.394E-21 & 8.19 & 42.89 & 6.2 & 32.5 & 4.99E+33 \\
5.2-2.6 & & S & 1.208E-21 & 8.39 & 43.93 & 8.4 & 44.0 & 7.93E+33 \\
5.4-1.2* & Milne 56 & C?(P) & 4.30E-21 & 3.39 & 34.48 & 4.5 & 45.8 &
3.06E+34 \\
5.9+3.1 & & S & 1.24E-21 & 7.52 & 43.73 & 7.5   & 43.6  & 8.02E+33 \\
6.1+1.2 & & F   & 7.72E-22      & & & & & \\
6.4-0.1* & W28  & C & 2.64E-20  & 1.18  & 16.53 & 2.5   & 34.9  & 
8.38E+34 \\
6.4+4.0 & & S   & 2.04E-22 & 6.56 & 59.14 & 6.5 & 58.6  & 2.38E+33 \\
7.0-0.1 & & S   & 1.67E-21 & 9.53 & 41.61 & 9.5 & 41.5  & 9.75E+33 \\
7.7-3.7 & 1814-24 & S & 3.42E-21 & 5.77 & 36.92 & 5.8   & 37.1  & 
1.60E+34 \\
8.7-5.0 & & S   & 9.80E-22 & 6.01 & 45.49 & 5.2 & 39.3  & 5.14E+33 \\
8.7-0.1 & W30 & S?(P) & 5.95E-21 & 2.31 & 30.24 & 3.5 & 45.8 & 4.24E+34 \\
9.8+0.6 & & S & 4.08E-21 & 10.11 & 35.24 & 12 & 41.8    & 2.43E+34 \\ 
10.0-0.3 & & ?  & 6.82E-21 & 12.35 & 28.61 & 12 & 27.8  & 1.8E+34 \\
11.2-0.3* & & C(XP) & 2.07E-19 & 6.28 & 7.19 & 5 & 5.7 & 2.38E+34 \\
11.4-0.1 & & S? & 1.41E-20 & 9.20 & 21.32 & 9.2 & 21.3  & 2.20E+34 \\
12.0-0.1 & & ?  & 1.08E-20 & 11.65 & 23.8 & 12  & 24.5  & 2.18E+34 \\
13.3-1.3 & & S? & ? & ? & ? & 3 & 46.0 & \\
13.5+0.2 & & S  & 2.63E-20 & 12.82 & 16.56 & 13 & 16.8  & 2.56E+34 \\
15.1-1.6 & & S  & 1.15E-21 & 5.68 & 44.29 & 5.7 & 44.5  & 7.72E+33 \\
15.9+0.2 & & S? & 2.15E-20 & 10.41 & 17.97 & 11 & 19.0  & 2.62E+34 \\
16.2-2.7 & & S  & 1.04E-21 & 9.11 & 45.03 & 8.8 & 43.5  & 6.69E+33 \\
16.7+0.1 & & C  & 2.82E-20 & 14.06 & 16.1 & 16  & 18.3  & 3.32E+34 \\ 
16.8-1.1 & & ?  & 4.18E-22 & 6.72 & 52.44 & 6.7 & 52.3  & 3.88E+33 \\

\end{tabular}
\end{flushleft}

\begin{flushleft}
\begin{tabular}{ccccccccc}
\multicolumn{9}{l}{\bf Table 1: Data of Galactic SNRs (Continued 1)} \\ 
\hline
l,b & Name & Type & $\Sigma$ & d$_{\Sigma - D}$ & D$_{\Sigma - D}$ &
d$_{ado}$ & D$_{ado}$ & L \\
& & & & (kpc) & (pc) & (kpc) & (pc) & erg/s \\
\hline
17.4-2.3 & & S & 1.25E-21 & 6.25 & 43.66 & 6.3 & 44.0  & 8.24E+33 \\
17.8-2.6 & & S  & 1.05E-21 & 6.45 & 45.01 & 6.4 & 44.7  & 7.08E+33 \\
18.8+0.3 & Kes 67 & S & 2.66E-20 & 4.14 & 16.5  & 8   & 31.8  &
9.13E+34 \\
18.9-1.1 & & C? & 5.11E-21 & 3.35 & 32.15 & 3.4 & 32.6  & 1.85E+34 \\
20.0-0.2 & & F  & 1.51E-20 & & & & & \\
21.5-0.9 & & C  & 6.27E-19 & 4.01 & 4.59 & 5.5  & 6.3   & 7.85E+33 \\
21.8-0.6 & Kes 69 & S & 2.60E-20 & 2.86 & 16.65 & 2.9   & 16.9  &
2.51E+34 \\
22.7-0.2 & & S? & 7.35E-21 & 3.67 & 27.76 & 3.7 & 28.0  & 1.95E+34 \\
23.3-0.3 & W41  & S & 1.45E-20  & 2.69  & 21.11 & 2.8   & 22.0  &   
2.37E+34 \\
23.6+0.3 & & ?  & 1.20E-20 & 7.81 & 22.73 & 8   & 23.3  & 2.21E+34 \\
24.7-0.6 & & S? & 5.35E-21 & 7.23 & 31.56 & 9   & 39.3  & 2.80E+34 \\
24.7+0.6 & & C? & 6.69E-21 & 4.68 & 28.84 & 5   & 30.8  & 2.16E+34 \\
27.4+0.0 & 4C-04.71 & S(AXP) & 5.64E-20 & 10.62 & 12.16 & 6.5 & 7.4  &
1.10E+34 \\
27.8+0.6 & & F  & 3.01E-21 & & & 3.2 & 36.1  & 1.33E+34 \\
28.6-0.1 & & S  & 3.86E-21 & 11.45 & 36 & 11.5  & 36.1  & 1.72E+34 \\
28.8+1.5 & & S? & ? & ? & ? & & & \\
29.6+0.1 & & S(AXP) & 9.03E-21 & 17.43 & 25.54 & 11 & 16.1  & 7.85E+33 \\
29.7-0.3 & Kes 75 & C(XP) & 1.67E-19 & 9.26 & 7.83 & 6.7  & 5.7   &
1.94E+34 \\
30.7-2.0 & & ?  & 2.94E-22 & 11.94 & 55.62 & 12 & 55.9  & 3.11E+33 \\
30.7+1.0 & & S? & 2.09E-21 & 6.63 & 40.09 & 6.6 & 39.9  & 1.13E+34 \\
31.5-0.6 & & S? & 9.29E-22 & 8.77 & 45.9 & 8.8  & 46.1  & 6.69E+33 \\ 
31.9+0.0* & 3C391 & S   & 1.03E-19 & 5.52 & 9.52 & 8.5  & 14.7  &
7.50E+34 \\
32.0-4.9 & 3C396.1 & S? & 9.20E-22 & 2.63 & 45.98 & 2.7 & 47.1  &
6.93E+33 \\
32.1-0.9 & & C? & ? & ? & ? & & & \\
32.8-0.1 & Kes 78 & S?  & 5.73E-21 & 6.21 & 30.7 & 6.3  & 31.1  &
1.89E+34 \\
33.2-0.6 & & S  & 1.63E-21 & 7.98 & 41.8 & 8 & 41.9 & 9.68E+33 \\
33.6+0.1 & Kes 79 & S   & 3.31E-20 & 5.19 & 15.09 & 7 & 20.3  &
4.66E+34 \\
34.7-0.4* & W44 & C(P) & 3.66E-20  & 1.62  & 14.49 & 2.8   & 25.0  &
7.79E+34 \\
36.6-0.7 & & S? & ? & ? & ? & & & \\
36.6+2.6 & & S & 4.77E-22 & 11.86 & 51.31 & 11.6  & 50.2  &
4.07E+33 \\
39.2-0.3 & 3C396 & S & 5.64E-20  & 6.04  & 12.16 & 7.7     & 15.5  &
4.61E+34 \\
39.7-2.0 & W50  & ? & 1.78E-21  & 1.67  & 41.19 & 5.0   & 124.0  &
9.17E+34 \\
40.5-0.5 & & S  & 3.42E-21 & 5.77 & 36.92 & 5.7 & 36.5  & 1.55E+34 \\

\end{tabular}
\end{flushleft}

\begin{tabular}{ccccccccc}
\multicolumn{9}{l}{\bf Table 1: Data of Galactic SNRs (Continued 2)} \\ 
\hline
l,b & Name & Type & $\Sigma$ & d$_{\Sigma - D}$ & D$_{\Sigma - D}$ &
d$_{ado}$ & D$_{ado}$ & L \\
& & & & (kpc) & (pc) & (kpc) & (pc) & erg/s \\
\hline
41.1-0.3 & 3C397 & S  & 2.94E-19 & 6.38 & 6.23 & 6.4  & 6.2   & 3.90E+34 \\
42.8+0.6 & & S  & 7.84E-22 & 6.76 & 47.22 & 6   & 41.9  & 4.67E+33 \\
43.3-0.2* & W49B & S & 4.77E-19 & 5.25 & 5.13  & 9     & 8.8   &
1.33E+35 \\
43.9+1.6 & & S? & 3.60E-22 & 3.08 & 53.78 & 3.1 & 54.1  & 3.57E+33 \\
45.7-0.4 & & S  & 1.31E-21 & 6.77 & 43.36 & 6.7 & 42.9  & 8.15E+33 \\
46.8-0.3 & HC30 & S & 9.53E-21 & 5.77 & 24.98   & 7.0   & 30.3  &
2.97E+34 \\
49.2-0.7 & W51  & S? & 2.68E-20 & 1.88  & 16.45 & 4.0   & 34.9  &
1.11E+35 \\
53.6-2.2 & 3C400.2 & S  & 1.30E-21 & 4.91 & 43.38 & 4.5 & 39.8  &
7.00E+33 \\
54.1+0.3 & & F & 3.34E-20 & & & 10 & 4.9 & 2.16E+33 \\
54.4-0.3* & HC 40 & S   & 2.63E-21 & 3.31 & 38.57 & 3.3 & 38.4  &
1.32E+34 \\
55.0+0.3 & & S  & 2.51E-22 & 11.33 & 57.11 & 11.3 & 56.9 & 2.76E+33 \\
55.7+3.4 & & S  & 3.98E-22 & 7.91 & 52.87 & 7   & 46.8  & 2.97E+33 \\
57.2+0.8 & 4C21.53 & S? & 1.88E-21 & 11.70 & 40.8 & 11.7 & 40.8 &
1.07E+34 \\
59.5+0.1 & & S  & 1.81E-20 & 13.17 & 19.29 & 11 & 16.1  & 1.57E+34 \\
59.8+1.2 & & ?  & 7.53E-22 & 9.14 & 47.54 & 9.1 & 47.3  & 5.73E+33 \\
63.7+1.1 & & F  & 4.23E-21 & & & & & \\
65.1+0.6 & & S  & 2.01E-22 & 3.04 & 59.28 & 3   & 58.5  & 2.33E+33 \\
65.3+5.7 & & S? & 1.05E-22 & 0.83 & 66.03 & 0.8 & 63.5  & 1.44E+33 \\
65.7+1.2 & DA 495 & ?   & 2.37E-21 & 7.50 & 39.26 & 7.5 & 39.3  &
1.24E+34 \\
67.7+1.8 & & S  & 2.60E-21 & 14.83 & 38.65 & 14 & 36.5  & 1.19E+34 \\
68.6-1.2 & & ?  & 1.51E-22 & 8.08 & 62.2 & 8    & 61.6  & 1.94E+33 \\
69.0+2.7* & CTB 80 & ?(P) & 2.82E-21 & 1.64 & 38.13 & 2   & 46.5  &
2.08E+34 \\
69.7+1.0 & & S  & 9.41E-22 & 9.83 & 45.8 & 9.5 & 44.2   & 6.24E+33 \\
73.9+0.9 & & S? & 2.80E-21 & 5.96 & 38.18 & 5   & 32  & 9.72E+33 \\
74.0-8.5* & Cygnus L. & S & 8.59E-22 & 0.83 & 46.5 & 0.8 & 44.6 &
5.81E+33 \\
74.9+1.2 & CTB 87 & F & 2.82E-20 & & & 7.9   & 15.9  & 2.43E+34 \\
76.9+1.0 & & ?  & 2.79E-21 & 12.61 & 38.21 & 12.6 & 38.2 & 1.37E+34 \\
78.2+2.1* & DR4 & S(D) & 1.42E-20 & 1.22  & 21.25 & 1.5   & 26.2  &  
3.31E+34 \\
82.2+5.3 & W63 & S & 2.92E-21 & 1.66 & 37.9 & 1.7       & 38.9  &
1.50E+34 \\
84.2-0.8 & & S  & 5.17E-21 & 6.15 & 32  & 5 & 26      & 1.19E+34 \\  
84.9+0.5 & & S  & 3.34E-21 & 21.05 & 37.06 & 10 & 17.6  & 3.46E+33 \\
85.4+0.7 & & S  & ? & ? & ? & & & \\
85.9-0.6 & & S  & ? & ? & ? & & & \\

\end{tabular}

\begin{flushleft}
\begin{tabular}{ccccccccc}
\multicolumn{9}{l}{\bf Table 1: Data of Galactic SNRs (Continued3)} 
\\ \hline
l,b & Name & Type & $\Sigma$ & d$_{\Sigma - D}$ & D$_{\Sigma - D}$ &
d$_{ado}$ & D$_{ado}$ & L \\
& & & & (kpc) & (pc) & (kpc) & (pc) & erg/s \\
\hline
89.0+4.7* & HB21 & S   & 3.07E-21 & 1.24 & 37.6 & 0.9  & 27.2  & 7.70E+33 \\
93.3+6.9* & DA 530 & S & 2.51E-21 & 5.75 & 38.89 & 3.8 & 25.7  &
5.62E+33 \\
93.7-0.2 & DA 551 & S   & 1.53E-21 & 1.82 & 42.24 & 1.6 & 37.2  &
7.19E+33 \\
94.0+1.0 & 3C434.1 & S  & 3.01E-21 & 4.73 & 37.72 & 4.7 & 37.5  &
1.43E+34 \\
106.3+2.7 & & ?(P) & 6.27E-22 & 4.44 & 49.01 & 5.5 & 60.7 & 7.85E+33 \\
109.1-1.0* & CTB 109 & S(AXP) & 3.84E-21 & 4.43 & 36.1 & 5 & 40.7 &
2.16E+34 \\
111.7-2.1 & Cas A & S(D) & 1.64E-17 & 0.84 & 1.22 & 3 & 4.4 & 1.06E+36 
\\   
114.3+0.3* & & S(P) & 1.82E-22 & 2.94 & 60.23 & 2.8 & 57.3 & 2.03E+33 \\
116.5+1.1 & & S & 3.45E-22 & 2.69 & 54.16 & 3.5 & 70.5  & 5.82E+33 \\
116.9+0.2* & CTB 1 & S  & 1.17E-21 & 4.46 & 44.15 & 3.5 & 34.6  &
4.77E+33 \\
117.7+0.6 & & (D) & & & & 3 & & \\
119.5+10.2* & CTA 1 & S(D) & 6.69E-22 & 1.85 & 48.49 & 1.4 & 36.7 &
3.05E+33 \\
120.1+1.4* & Tycho & S  & 1.32E-19 & 3.73 & 8.63 & 3.3  & 7.6   &  
2.64E+34 \\
126.2+1.6 & & S? & 2.15E-22 & 2.88 & 58.6 & 2.5 & 50.9  & 1.89E+33 \\
127.1+0.5 & R5  & S & 9.66E-22  & 3.48  & 45.6  & 2.5   & 32.7  &
3.51E+33 \\
130.7+3.1 & SN 1181 & F(D) & 1.10E-19 & & & 3.2 & 5.8   & 1.46E+34 \\
132.7+1.3* & HB3 & S & 1.06E-21 & 1.93 & 44.91 & 2.3 & 53.5  &
1.03E+34 \\
156.2+5.7* & & S & 6.22E-23 & 2.25 & 72.08 & 2.3 & 73.6  & 1.14E+33 \\
160.9+2.6 & HB9 & S & 9.85E-22  & 1.21  & 45.45 & 1.2   & 45.2  &
6.85E+33 \\
166.0+4.3* & VRO & S & 5.47E-22 & 3.93 & 50.14 & 3.8   & 48.5 &
4.37E+33 \\
166.2+2.5 & OA 184 & S  & 2.63E-22 & 2.45 & 56.67 & 2.5 & 57.7  &
2.97E+33 \\
179.0+2.6 & & S? & 2.15E-22 & 2.88 & 58.6 & 2.9 & 59.0  & 2.55E+33 \\
180.0-1.7* & S 147 & S  & 3.02E-22 & 1.06 & 55.37 & 1   & 52.4  &
2.81E+33 \\
182.4+4.3 & & S & 7.22E-23 & 4.83 & 70.3 & 3.5  & 50.9  & 6.35E+32 \\ 
184.6-5.8 & Crab & F(P) & 4.47E-18 & &  & 2     & 3.5   &
1.80E+35 \\
189.1+3.0* & IC 443 & C & 1.19E-20 & 1.75 & 22.84 & 1.5 & 19.6  &
1.56E+34 \\
192.8-1.1 & PKS 0607 & S & 4.95E-22 & 2.25 & 50.99 & 2.3 & 52.2 &
4.57E+33 \\
205.5+0.5* & Monoceros & S & 4.98E-22 & 0.80 & 50.94 & 1 & 64.0 &
6.92E+33 \\
206.9+2.3 & PKS 0646 & S? & 3.76E-22 & 3.74 & 53.37 & 3.4 & 48.5 &
3.00E+33
\\
260.4-3.4* & Puppis A & S(D) & 6.52E-21 & 1.83 & 29.13 & 2 & 31.9  &
2.25E+34 \\
261.9+5.5 & & S & 1.25E-21 & 4.33 & 43.66 & 3.3 & 33.3 & 4.71E+33 \\
263.9-3.3* & Vela & C(P) & 4.05E-21 & 0.48 & 35.33 & 0.45 & 33.4 &
1.53E+34 \\
266.2-1.2 & & S(D) & 5.23E-22 & 1.45 & 50.53 & 1.0 & 34.9  & 2.16E+33 \\
272.2-3.2 & & S? & 2.68E-22 & 12.94 & 56.5 & 9 & 39.3 & 1.40E+33 \\

\end{tabular}
\end{flushleft}

\begin{flushleft}
\begin{tabular}{ccccccccc}
\multicolumn{9}{l}{\bf Table 1: Data of Galactic SNRs (Continued 4)} 
\\ \hline
l,b & Name & Type & $\Sigma$ & d$_{\Sigma - D}$ & D$_{\Sigma - D}$ &
d$_{ado}$ & D$_{ado}$ & L \\
& & & & (kpc) & (pc) & (kpc) & (pc) & erg/s \\
\hline
279.0+1.1 & & S & 5.00E-22 & 1.84 & 50.9 & 1.8  & 49.7  &
4.20E+33 \\
284.3-1.8 & MSH10-53 & S(P) & 2.87E-21 & 5.44 & 38.01 & 5.2 & 36.3 &
1.29E+34 \\
286.5-1.2 & & S? & 1.35E-21 & 11.85 & 43.12 & 10 & 36.4 & 6.05E+33 \\
289.7-0.3 & & S & 3.70E-21 & 7.93 & 36.64 & 7.9 & 36.5  & 1.67E+34 \\
290.1-0.8 & & S & 2.38E-20 & 3.64 & 17.26 & 5.5 & 26.1  & 5.48E+34 \\
291.0-0.1 & & C & 1.23E-20 & 5.55 & 22.5 & 5.5  & 22.3  & 2.09E+34 \\
292.0+1.8 & MSH11-54 & C & 2.35E-20 & 6.09 & 17.33 & 5.0 & 14.3  &
1.62E+34 \\
292.2-0.5 & & S(P) & 3.51E-21 & 7.30 & 36.76 & 7.5 & 37.8  & 1.70E+34 \\
293.8+0.6 & & C & 1.88E-21 & 7.01 & 40.8 & 6.9 & 40.1   & 1.03E+34 \\
294.1-0.0 & & S & $>$1.88E-22 & ? & $<$59.92 & 4 & & $>$1.38E+33 \\
296.1-0.5 & & S & 1.30E-21 & 4.91 & 43.39 & 4.9 & 43.3  & 8.30E+33 \\
296.5+10* & PKS1209 & S(D) & 1.23E-21 & 1.97 & 43.77 & 1.8 & 40.1 &
6.72E+33 \\
296.8-0.3 & 1156-62 & S & 4.84E-21 & 6.75 & 32.88 & 6.8 & 33.1 & 1.80E+34 \\
298.5-0.3 & & ? & 3.01E-20 & 10.71 & 15.68 & 11 & 16.1  & 2.62E+34 \\
298.6-0.0 & & S & 6.97E-21 & 9.36 & 28.36 & 9.3 & 28.2  & 1.87E+34 \\
299.2-2.9 & & S & 3.80E-22 & 13.00 & 53.29 & 12 & 49.2  & 3.11E+33 \\
299.6-0.5 & & S & 8.91E-22 & 12.22 & 46.22 & 12.2 & 46.1 & 6.43E+33 \\
301.4-1.0 & & S & 3.71E-22 & 6.30 & 53.49 & 6.3 & 53.5  & 3.60E+33 \\
302.3+0.7 & & S & 2.60E-21 & 7.82 & 38.64 & 7.8 & 38.6  & 1.31E+34 \\
304.6+0.1 & Kes 17 & S & 3.29E-20 & 6.53 & 15.13 & 6.5 & 15.1 & 2.56E+34 \\
308.1-0.7 & & S & 1.07E-21 & 11.86 & 44.84 & 11 & 41.6 & 6.28E+33 \\ 
308.8-0.1 & & C?(P) & 3.76E-21 & 5.11 & 36.4 & 8 & 57.0  & 4.15E+34 \\
309.2-0.6 & & S & 5.85E-21 & 7.79 & 30.44 & 6 & 23.5  & 1.09E+34 \\
309.8+0.0 & & S & 5.39E-21 & 4.97 & 31.48 & 5 & 31.7 & 1.84E+34 \\   
310.6-0.3 & Kes 20B & S & 1.18E-20 & 9.91 & 22.95 & 9.9 & 22.9 & 2.12E+34 \\
310.8-0.4 & Kes 20A & S & 6.27E-21 & 8.49 & 29.6 & 8.5  & 29.6  &
1.87E+34 \\
311.5-0.3 & & S & 1.81E-20 & 13.17 & 19.29 & 12 & 17.6  & 1.87E+34 \\
312.4-0.4 & & S & 4.69E-21 & 3.01 & 33.29 & 3 & 33.2 & 1.75E+34 \\   
315.4-2.3* & RCW 86 & S & 4.18E-21 & 2.86 & 34.88 & 2.7 & 33.0 &
1.54E+34 \\
315.4-0.3 & & ? & 3.86E-21 & 7.02 & 36.03 & 7 & 35.9 & 1.69E+34 \\
315.9-0.0 & & S & 3.44E-22 & 9.96 & 54.18 & 10 & 54.4 & 3.46E+33 \\
316.3-0.0 & MSH14-57 & S & 7.41E-21 & 4.72 & 27.66 & 4.7 & 27.5 &   
1.91E+34 \\
317.3-0.2 & & S & 5.85E-21 & 9.51 & 30.45 & 9.5 & 30.4  & 1.83E+34 \\

\end{tabular}
\end{flushleft}

\begin{flushleft}
\begin{tabular}{ccccccccc}
\multicolumn{9}{l}{\bf Table 1: Data of Galactic SNRs (Continued 5)} 
\\ \hline
l,b & Name & Type & $\Sigma$ & d$_{\Sigma - D}$ & D$_{\Sigma - D}$ &
d$_{ado}$ & D$_{ado}$ & L \\
& & & & (kpc) & (pc) & (kpc) & (pc) & erg/s \\
\hline
318.2+0.1 & & S & $>$4.19E-22 & ? & $<$52.42 & 4 & & \\
318.9+0.4 & & C & 1.43E-21 & 7.16 & 42.69 & 7.2 & 42.9 & 8.96E+33 \\
320.4-1.2* & RCW 89 & C(P) & 7.37E-21 & 2.72 & 27.72 & 4.2 & 42.8 &
4.58E+34 \\ 320.6-1.6 & & S & ? & ? & ? & & & \\
321.9-1.1 & & S & $>$6.53E-22 & ? & $<$48.69 & 5 & & $>$3.67E+33 \\
321.9-0.3 & & S & 2.74E-21 & 4.93 & 38.31 & 4.9 & 38.1  & 1.35E+34 \\
322.5-0.1 & & C & 1.00E-21 & 10.38 & 45.31 & 10.3 & 45.0 & 6.88E+33 \\
323.5+0.1 & & S & 2.67E-21 & 10.17 & 38.48 & 10 & 37.8 & 1.30E+34 \\
326.3-1.8 & MSH15-56 & C & 1.51E-20 & 1.88 & 20.73 & 2.0 & 22.1 &
2.5E+34 \\
327.1-1.1 & & C & 3.25E-21 & 7.11 & 37.24 & 6.5 & 34.1  & 1.28E+34 \\
327.4+0.4 & Kes 27 & S  & 1.02E-20 & 3.97 & 24.27 & 5.0 & 30.5  &   
3.24E+34 \\
327.4+1.0 & & S & 1.46E-21 & 10.46 & 42.57 & 10.4 & 42.3 & 8.88E+33 \\
327.6+14.6* & SN 1006 & S & 3.18E-21 & 4.28 & 37.38 & 2 & 17.5  &  
3.29E+33 \\
328.4+0.2 & MSH15-57 & F & 9.03E-20 & & & 17.4 & 25.5 &
1.96E+35 \\
329.7+0.4 & & S & $>$3.88E-21 & ? & $<$35.96 & 3.3 & & $>$1.6E+34 \\
330.0+15.0 & Lupus L. & S & 1.63E-21 & 0.80 & 41.8 & 0.8 & 41.9 &
9.69E+33 \\
330.2+1.0 & & S? & 6.22E-21 & 9.28 & 29.7 & 9.3 & 29.8 & 1.87E+34 \\
332.0+0.2 & & S & 8.36E-21 & 7.56 & 26.34 & 7.5 & 26.1 & 1.95E+34 \\
332.4-0.4* & RCW 103 & S & 4.21E-20 & 4.70 & 13.69 & 3.7 & 10.8 &
1.66E+34 \\
332.4+0.1 & Kes 32 & S(D) & 1.74E-20 & 4.48 & 19.58 & 4 & 17.5 &
1.80E+34 \\
335.2+0.1 & & S & 5.46E-21 & 5.12 & 31.3 & 5.1 & 31.1 & 1.80E+34 \\
336.7+0.5 & & S & 6.45E-21 & 8.52 & 29.26 & 8.5 & 29.2 & 1.87E+34 \\
337.0-0.1 & CTB 33 & S & 1.00E-19 & 19.73 & 9.63 & 11 & 5.4   &
7.86E+33 \\
337.2-0.7 & & S & 8.36E-21 & 14.96 & 26.34 & 12 & 21.1 & 1.25E+34 \\
337.3+1.0 & Kes 40 & S & 1.34E-20 & 5.58 & 21.78 & 5.6 & 21.9 & 2.17E+34 \\
337.8-0.1 & Kes 41 & S & 5.02E-20 & 5.99 & 12.75 & 8 & 17.1 & 4.98E+34 \\
338.1+0.4 & & S & 2.68E-21 & 8.81 & 38.47 & 8.8 & 38.4 & 1.34E+34 \\
338.3-0.0 & & S & 1.65E-20 & 8.65 & 20.03 & 8.6 & 19.9  & 2.24E+34 \\
338.5+0.1 & & ? & 2.23E-20 & 6.79 & 17.71 & 6.8 & 17.7  & 2.40E+34 \\
340.4+0.4 & & S & 1.08E-20 & 9.75 & 23.8 & 9.7 & 23.7 & 2.03E+34 \\
340.6+0.3 & & S & 2.09E-20 & 10.33 & 18.18 & 10.3 & 18.1 & 2.29E+34 \\
341.2+0.9 & & C(P) & 6.41E-22 & 8.94 & 48.83 & 6.8 & 37.1 & 3.00E+33 \\
341.9-0.3 & & S & 7.68E-21 & 13.35 & 27.27 & 13.3 & 27.2 & 1.91E+34 \\

\end{tabular}
\end{flushleft}

\begin{flushleft}
\begin{tabular}{ccccccccc}
\multicolumn{9}{l}{\bf Table 1: Data of Galactic SNRs (Continued 6)} 
\\ \hline
l,b & Name & Type & $\Sigma$ & d$_{\Sigma - D}$ & D$_{\Sigma - D}$ &
d$_{ado}$ & D$_{ado}$ & L \\
& & & & (kpc) & (pc) & (kpc) & (pc) & erg/s \\
\hline
342.0-0.2 & & S & 4.88E-21 & 10.82 & 32.77 & 10.8 & 32.7 & 1.76E+34 \\
342.1+0.9 & & S & 8.36E-22 & 16.91 & 46.71 & 16.9 & 46.7 & 6.17E+33 \\
343.0-6.0 & & S & ? & ? & ? & & & \\
343.1-2.3 & & C? & 1.18E-21 & 4.74 & 44.13 & 4.7 & 43.8 & 7.64E+33 \\
343.1-0.7 & & S & 2.07E-21 & 5.80 & 40.15 & 5.8 & 40.2  & 1.13E+34 \\
344.7-0.1 & & C? & 3.76E-21 & 12.51 & 36.4 & 12.5 & 36.4 & 1.69E+34 \\
345.7-0.2 & & S & 2.51E-21 & 22.09 & 38.89 & 18 & 31.7  & 8.40E+33 \\
346.6-0.2 & & S & 1.88E-20 & 8.19 & 18.97 & 8.2 & 19.0  & 2.33E+34 \\
347.3-0.5 & & S? & ? & ? & ? & 6.0 & & \\
348.5-0.0 & & S? & 1.51E-20 & 7.14 & 20.77 & 7.1 & 20.7 & 2.18E+34 \\
348.5+0.1 & CTB 37A & S & 4.82E-20 & 2.97 & 12.97 & 6.0 & 26.2 &
1.12E+35 \\
348.7+0.3 & CTB 37B & S & 1.35E-20 & 4.38 & 21.67 & 7.0 & 34.5  &
5.52E+34 \\
349.2-0.1 & & S & 3.90E-21 & 16.85 & 35.87 & 16 & 34.1  & 1.55E+34 \\
349.7+0.2 & & S & 6.02E-19 & 6.75 & 4.66 & 12.0  & 8.2 & 1.24E+35 \\  
350.0-2.0 & & S & 1.93E-21 & 3.10 & 40.62 & 3.1 & 40.6  & 1.08E+34 \\
351.2+0.1 & & C? & 1.54E-20 & 10.08 & 20.6 & 10.1 & 20.6 & 2.20E+34 \\
351.7+0.8 & & S & 5.97E-21 & 6.54 & 30.19 & 6.5 & 30.0 & 1.83E+34 \\
351.9-0.9 & & S & 2.51E-21 & 12.83 & 38.89 & 12.8 & 38.8 & 1.27E+34 \\
352.7-0.1 & & S & 1.25E-20 & 11.10 & 22.36 & 11 & 22.1  & 2.09E+34 \\
353.9-2.0 & & S & 8.91E-22 & 12.22 & 46.22 & 12 & 45.4 & 6.23E+33 \\
354.1+0.1 & & C? & ? & ? & ? & & & \\
354.8-0.8 & & S & 1.17E-21 & 8.00 & 44.18 & 8 & 44.2 & 7.75E+33 \\
355.6-0.0 & & S & 9.41E-21 & 12.48 & 25.12 & 12 & 24.2 & 1.87E+34 \\ 
355.9-2.5 & & S & 7.12E-21 & 7.43 & 28.11 & 7.4 & 28.0 & 1.89E+34 \\  
356.2+4.5 & & S & 9.63E-22 & 6.27 & 45.62 & 6 & 43.6 & 6.23E+33 \\   
356.3-0.3 & & S & 5.86E-21 & 11.88 & 30.41 & 11.8 & 30.2 & 1.81E+34 \\
356.3-1.5 & & S & 1.51E-21 & 8.40 & 42.35 & 8.4 & 42.3 & 9.15E+33 \\
357.7-0.1 & MSH 17-39 & ? & 2.32E-19 & 4.82 & 6.86 & 7.0 & 9.9 &
7.85E+34 \\
357.7+0.3 & & S & 2.61E-21 & 5.53 & 38.62 & 5.5 & 38.4 & 1.31E+34 \\
358.0+3.8 & & S & 1.56E-22 & 5.59 & 61.8 & 5.2  & 57.5  & 1.75E+33 \\
359.0-0.9 & & S & 6.54E-21 & 4.35 & 29.09 & 4.4 & 29.4  & 1.92E+34 \\
359.1-0.5 & & S & 3.66E-21 & 5.23 & 36.51 & 5.2 & 36.3 & 1.64E+34 \\ 
359.1+0.9 & & S & 5.70E-21 & 9.19 & 30.76 & 9.2 & 30.8  & 1.83E+34 \\ 
\hline

\end{tabular}
\end{flushleft}

\clearpage
\begin{figure}[t]
\vspace{3cm}
\includegraphics{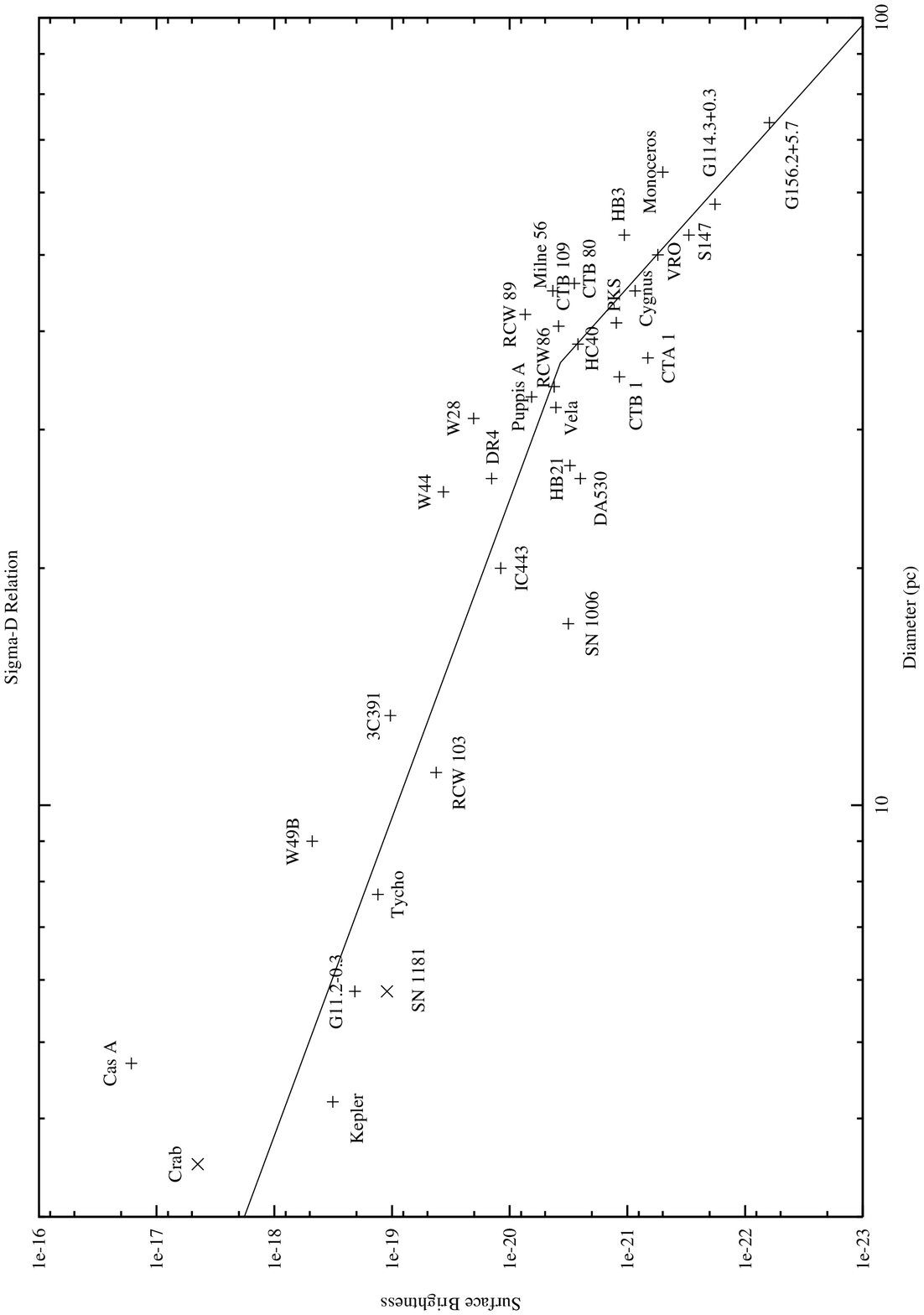}
\end{figure}

\clearpage
\begin{figure}[t]
\vspace{3cm}
\includegraphics{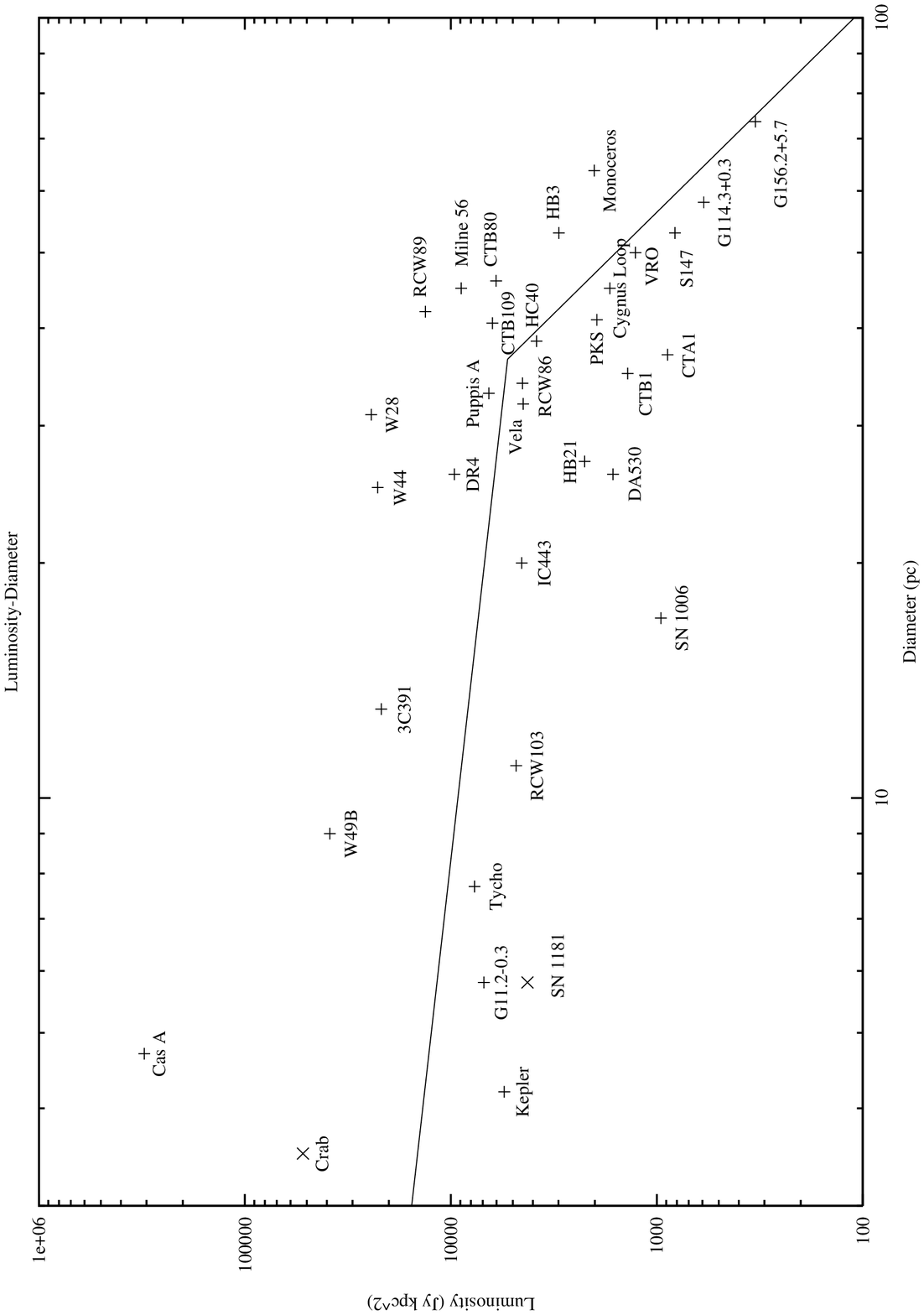}
\end{figure}

\end{document}